\documentclass[a4paper,fleqn,usenatbib,useAMS]{mnras}
\usepackage{color}
\usepackage{amsmath}	
\usepackage{amssymb}	
\usepackage{multicol}        
\usepackage{bm}		
\usepackage{textcomp,amssymb}
\usepackage{leftidx}
\usepackage{array}
\usepackage{supertabular}
\usepackage{hhline} 
\usepackage{multirow}
\usepackage{balance}
\usepackage{times}
\usepackage{graphicx}
\usepackage{calrsfs}
\DeclareMathAlphabet{\pazocal}{OMS}{zplm}{m}{n}
\usepackage{soul}
\usepackage{bm}
\usepackage{balance}
\usepackage{hyperref} 
\hypersetup{colorlinks=true, linkcolor=blue, citecolor=blue, filecolor=blue, urlcolor=black}

\def \apj {{\it ApJ}}
\def \apjl {{\it ApJ Let.}}
\def \apjs {{\it ApJ Sup.}}
\def \pasj {{\it PASJ}}

\def \aap {{\it A\&A}}

\def \mnras {{\it MNRAS}}
\def \nat {{\it Nature}}

\definecolor{lblue}{rgb}{0.1,0.7,1.}
\definecolor{grey}{rgb}{0.75,0.75,0.75}
\definecolor{Orange}{rgb}{1.0,0.5,0.15}
\definecolor{brown}{rgb}{0.7,0.25,0.0}
\definecolor{pink}{rgb}{1.0,0.5,0.5}
\definecolor{darkerred}{rgb}{0.8,0,0}
\definecolor{darkerblue}{rgb}{0,0,0.8}
\definecolor{Blue}{rgb}{0,0.08,0.65}
\definecolor{Red}{rgb}{0.65,0.08,0.05}
\definecolor{Green}{rgb}{0.35,0.45,0.25}

\begin{document}

\author[E. Darragh-Ford, C. Laigle, G. Gozaliasl et al.]{
\parbox[t]{\textwidth}{E.~Darragh~Ford$^{1,2}$\thanks{elise.darragh-ford.17@alumni.ucl.ac.uk}, C.~Laigle$^{2}$\thanks{clotilde.laigle@physics.ox.ac.uk}, G.~Gozaliasl$^{3,4,5}$, C.~Pichon$^{6,7,8}$,  J.~Devriendt$^{2}$, A.~Slyz$^{2}$,\\ S.~Arnouts$^{9}$, Y.~Dubois$^{6}$, A.~Finoguenov$^{3,4}$, R.~Griffiths$^{10}$, K.~Kraljic$^{8}$, H.~Pan$^{2,11}$, S.~Peirani$^{12}$, F.~Sarron$^{6}$
}
\vspace*{6pt} \\ 
$^{1}$ Physics, University of Chicago, 5640 South Ellis Avenue, ERC 569, Chicago, IL 60637, USA\\
$^{2}$ Sub-department of Astrophysics, University of Oxford, Keble Road, Oxford OX1 3RH\\
$^{3}$ Finnish centre for Astronomy with ESO (FINCA), Quantum, Vesilinnantie 5, University of Turku, FI-20014 Turku, Finland;\\
$^{4}$ Department of Physics, University of Helsinki, PO Box 64, FI-00014 Helsinki, Finland \\
$^{5}$Helsinki Institute of Physics, University of Helsinki, P.O. Box 64, FI-00014, Helsinki, Finland\\
$^{6}$ Sorbonne Universit{\'e}s, UPMC Univ Paris 6 et CNRS, UMR 7095, Institut d'Astrophysique de Paris, 98 bis bd Arago, 75014 Paris \\
$^{7}$Institute for Astronomy, University of Edinburgh, Royal Observatory, Blackford Hill, Edinburgh, EH9 3HJ, United Kingdom\\
$^{8}$ Korea Institute for Advanced Study (KIAS), 85 Hoegiro, Dongdaemun-gu, Seoul, 02455, Republic of Korea \\
$^{9}$ Aix Marseille Univ, CNRS, CNES, LAM, Marseille, France \\
$^{10}$ Dept. of Physics \& Astronomy, University of Hawaii at Hilo, 200 W. Kawili St., Hilo, Hi 96720, USA \\
$^{11}$ University of Chinese Academy of Sciences, Beijing 100049, China \\
$^{12}$ Universit\'e C\^ote d'Azur, Observatoire de la C\^ote d'Azur, CNRS, Laboratoire Lagrange, Bd de l'Observatoire, CS 34229, F-06304 Nice Cedex 4, France
}
\date{Accepted . Received ; in original form }

\title[Group connectivity in COSMOS]{
Group connectivity in COSMOS: a tracer of mass assembly history}

\maketitle
\begin{abstract}
Cosmic filaments are the channel through which galaxy groups assemble their mass. Cosmic connectivity, namely the number of filaments connected to a given group, is therefore expected to be an important ingredient in shaping group properties. \\
The local connectivity is measured in COSMOS around X-Ray detected groups between redshift~0.5 and~1.2. To this end, large-scale filaments are extracted  using the accurate photometric redshifts  of the {COSMOS2015} catalogue in two-dimensional slices of thickness 120~comoving~Mpc centred on the group's redshift. The link between connectivity, group mass and the properties of the brightest group galaxy (BGG) is investigated. The same measurement is carried out on mocks extracted from the lightcone of the hydrodynamical simulation {\sc Horizon-AGN} in order to control systematics. 
\\
More massive groups are on average more connected. At fixed group mass in low-mass groups, BGG mass is slightly enhanced at high connectivity, while in high mass groups BGG mass is lower at  higher connectivity. Groups with a star-forming BGG have on average a lower connectivity at given mass. From the analysis of the {\sc Horizon-AGN}  simulation, we postulate that different connectivities trace different paths of group mass assembly: at high group mass, groups with higher connectivity are more likely to have grown through a recent major merger, which might be in turn the reason for the quenching of the BGG. Future large-field photometric  surveys, such as {\emph{Euclid}} and LSST, will be able to confirm and extend these results by probing a wider mass range and a larger variety of environment. 
\end{abstract}


\section{Introduction}

In the local Universe, a large fraction of the stellar mass resides in galaxy groups and clusters. These environments  are the place of a wide variety of quenching processes, mostly depending on the group mass,  leading to different galaxy population contents \citep[e.g.][]{gobat15,treyer18}.  
These processes are first connected to the availability and physical state of the intra-cluster gas. The infalling cold gas can be gravitationally heated \citep[e.g.][]{birnboim03,keres05}, maintained hot via feedback from Active Galaxy Nuclei \citep[AGN,][]{dubois13}, and therefore  not available anymore for accretion and star formation. Other processes are related to interactions between gas and galaxies, such as ram-pressure stripping \citep{Gunn1972}, or between galaxies, including galaxy mergers, galaxy harassment \citep{mooreetal1996} or tidal interactions \citep{byrd90}. In addition, the infalling gas in the group or cluster can be diverted towards the central galaxy or its satellites, depending on their relative mass \citep[e.g.][]{simha09} and the relaxation state of the group, a process often leading to satellite quenching by starvation \citep[e.g.][]{voort17} due to tidal effect in the neighbourhood of the most massive galaxy.  

Groups are not isolated structures but they keep accreting matter --~including small galaxies, but also gas \citep{kauffmann10}~-- from the large-scale cosmic web they are connected to. Therefore, investigating the link between the large-scale cosmic web and group properties is an essential question for galaxy formation but also cosmology. 
On the one hand, the  mean number of filaments branching out from groups -- namely their local connectivity --  depends on the growth factor and therefore the Dark Energy equation of state. Precise measurement of this quantity  provides a topologically robust alternative  to constrain cosmology, as it can be shown to 
depend on moments  of the  hierarchy of the N-point correlation functions \citep{codis18}. The  disconnection of filaments with cosmic time is driven both by gravitational clustering and by dark energy which will stretch and disconnect neighbouring filaments through the increased expansion of voids.

On the other hand, the geometry and anisotropy  of the large-scale environment is connected to the nature, history and dynamics of  matter infall, knowledge of which, in turn, is  essential to shape the mass assembly of groups and clusters. Several works have already emphasized the link between the large-scale environment and the properties of groups and their central galaxies \citep[e.g.][]{scudder12,luparello15,zehavi18}, although some others conclude on the absence of correlation for high-mass halos \citep{jung14}.  The variety of definitions for the large-scale environment can be one of the reasons for these conflicting conclusions. On a similar note, the measured correlation between the quenching of the central galaxy and the fraction of quenched satellites in the group, introduced as 'galactic conformity' by \cite{weinmannetal2006}, and  either confirmed or debated since then \citep[e.g.][]{kauffman13,hearin16,hartley15,treyer18}, can be interpreted as an environmental process, in which the properties of both central and satellite galaxies of a given group depend on the mass assembly history of their host halo, which depends on the large-scale environment. 
 In fact, the geometry of  matter infall on groups being for the most part filamentary, this environmental question must be addressed primarily from the perspective of the cosmic web. 
Several questions are still unanswered: where do the groups sit in the cosmic web as a function of their mass? Can the filaments penetrate deeply into  halos and preferentially feed the central galaxies, or are the satellites bringing in all the accreted gas? 
What fraction of the mass load and the angular momentum is advected by 
how many filaments \citep[e.g.][]{pichon11, tillson11,danovichetal12,danovich15}? Can we understand cosmic connectivity as a function of the rareness of the nodes and prominence of filaments?
How is  (stellar or AGN) feedback impacted by the number of connected filaments?
Observationally, detecting filaments around groups is an essential first step to address these questions. 

This quest has therefore raised the interest of many and is now a growing field of investigation, with various methods of filament detections, including weak lensing \citep{dietrich05,jauzac12,gouin17}, stacking thermal Sunyaev-Zel'dovich detection \citep{bonjean18} or galaxy overdensity \citep{zhang13} between clusters pairs, X-Ray emission \citep{dietrich12,parekh17} or measurement of the local elongation spotted in the galaxy density distribution \citep{darvish15,durret16}.  Beyond trying to understand the process of mass assembly in the largest virialized structures of the Universe, the motivation for hunting matter in filaments is also to create a comprehensive census of the baryonic matter in the Universe. 
Although these first detections are  encouraging, a more systematic study is needed in order to draw statistically significant conclusions on the impact of the large-scale environment on galaxy group  properties.  These studies require a complete catalogue of accurate galaxy redshifts in order to precisely identify filaments, and a volume large enough to host both a significant number of massive structures and a large variety of large-scale environments. Large spectroscopic surveys are promising for this purpose, but are for now limited to low redshift \citep[see e.g.][for a study in the SDSS]{poudel18}. An alternative is to rely on photometric data. As shown in \cite{Laigle2018}, filaments can be reliably extracted from  photometric redshifts by relying on two-dimensional slices, the thickness of which is calibrated based on the typical redshift accuracy. As an example, extracting filaments from photometric redshifts around massive clusters has been already successfully done in the CFHTLS T0007 data \citep{sarron19} at low redshift ($0.15<z<0.70$).

In this work, we rely on the wealth of photometric data from the COSMOS field to perform the filament extraction around intermediate-mass groups at higher redshift, and we make use of a robust group catalogue which has been extracted from X-Ray photometry \citep{finoguenov07,gozaliasl19}. The correlation between group connectivity, group mass  and the BGG properties is first quantified both in the observations and the corresponding mocks. In the second step,  the {\sc Horizon} suite \citep{dubois14} is used to interpret the observational results.
The paper is organised as follows. Section~\ref{Sec:dataset} describes the observed and simulated datasets and
 the tools used to extract the skeleton.  Section~\ref{Sec:Results} presents the measurements both from  the observed and  simulated catalogues, and an extensive assessment of the robustness of the results. Section~\ref{Sec:Discussion} provides an interpretation of the observational results based on the {\sc Horizon-AGN} simulation. Section~\ref{Sec:Conclusion} summarizes the results and outlines future works. 
Appendix~\ref{sec:counts} gives more details on the connectivity measurement in the simulation and in COSMOS. 
We use a standard $\Lambda$CDM cosmology with a Hubble constant $H_{0}=70.4$ km$\cdot$s$^{-1}\cdot {\rm Mpc}^{-1}$, total matter density $\Omega_{\rm m}=0.272$ and dark energy density $\Omega_{\Lambda}=0.728$. Unless specified otherwise, errorbars are the errors on the mean derived from bootstrap resampling. 
%
%
\section{Dataset and extraction methods}
\label{Sec:dataset}
Let us first present briefly the observational dataset, the simulations  and the ridge extraction tools used to quantify the connectivity of groups, and the robustness of this extraction with respect to photometric redshift uncertainty.
 
\subsection{The COSMOS dataset}
\label{Sec:COSMOS}
The analysis presented here is based on the COSMOS dataset \citep{scovilleetal2007}. The cosmic web is extracted using the photometric redshift of the COSMOS2015 catalogue  \citep{Laigle2016}, which is also used for deriving galaxy masses. Groups are extracted from the  X-Ray photometry as described in \cite{finoguenov07} and \cite{gozaliasl19}. The observed catalogue of groups, their associated BGG and filaments in COSMOS is called $\mathcal{C}_{\rm cosmos}$ in the following analysis.

 \begin{table*}
 \begin{center}
 \def\arraystretch{1.15}
 \begin{tabular}{|c | c c c c c|}
 \hline
   \textbf{Name} & \textbf{Data} & \textbf{Group selection} & \textbf{$z$ and mass ranges} & \textbf{CW extraction and persistence} & \textbf{comments} \\ \hline
   $\mathcal{C}_{\rm cosmos}$ & COSMOS field & X-Ray & 0.5$<z<$1.2 & in 2D slices (120Mpc), 1.5$\sigma$ & photo-$z$ and  \\
   & 1.38 deg$^{2}$ & 86 groups & 13.38 $< \log M_{\rm group}/{\rm M}_{\odot}$ & from galaxies, $M_{*}>10^{10}{\rm M}_{\odot}$ &  masses \\\hline
   $\mathcal{C}_{\rm Hzagn\,2D}^{\rm phot}$ & {\sc Horizon-AGN} ligthcone & {\sc AdaptaHOP} & 0.5$<z<$1.2 & in 2D slices (120Mpc), 1.5$\sigma$ & photo-$z$ and \\
   & 1 deg$^{2}$ & 76 groups & 13.3 $< \log M_{\rm group}/{\rm M}_{\odot}$ & from galaxies, $M_{*}>10^{10}{\rm M}_{\odot}$ &   masses \\\hline
   $\mathcal{C}_{\rm Hzagn\,2D}^{\rm true}$ & {\sc Horizon-AGN} ligthcone & {\sc AdaptaHOP} & 0.5$<z<$1.2 & in 2D slices (120Mpc), 1.5$\sigma$ & intrinsic-$z$ \\
   & 1 deg$^{2}$ & 76 groups & 13.3 $< \log M_{\rm group}/{\rm M}_{\odot}$ & from galaxies, $M_{*}>10^{10}{\rm M}_{\odot}$ &  and masses \\\hline
   $\mathcal{C}_{\rm Hzagn\,3D}$ & {\sc Horizon-AGN} snapshots & {\sc AdaptaHOP} & 3 snapshots, $z=0.63,\,0.81,\,1.03$ & in 3D, 5$\sigma$ & intrinsic-$z$ \\
   & $3\times 100^{3}$ (Mpc$/h$)$^{3}$ & 1115 groups & 13.0 $< \log M_{\rm group}/{\rm M}_{\odot}$ & from all DM halos &  and masses \\\hline
 \end{tabular}
 \end{center}
 \caption{A summary of all the catalogues used  in this study, the data selection, the group and CW extraction methods and the persistence thresholds used in {\sc DisPerSE}. }
 \label{Tab:conf}
 \end{table*}
\subsubsection{COSMOS2015 redshifts and stellar masses}
The COSMOS2015 catalogue provides apparent 
magnitudes in 30 bands from ultra-violet (UV) to infra-red (IR). The photometric data include the optical COSMOS-20 subaru survey \citep{capak2007,ilbert2009}, Subaru Suprime-Cam data \citep{taniguchietal2007,taniguchietal2015}, the $u^{*}$-band data from the Canada-Hawaii-France Telescope (CFHT/MegaCam), NIR photometry  from the UltraVISTA survey \citep[DR2,][]{mccracken12}, the $Y$ band from the Hyper Suprime-Cam at Subaru telescope \citep{miyazaki2012} and mid-IR data in the four IRAC channels (i.e.~in a wavelength range between $\sim3$ and $8\,\mu$m) from the SPLASH program (PI: Capak).
Galaxy photometry in optical and NIR has been extracted using {\sc SExtractor} \mbox{\citep{bertinetal96}} in dual image mode, using as the detection image a $\chi^{2}$~sum of the four NIR images of UltraVISTA DR2 and the $z^{++}$-band (taken with Subaru Suprime-Cam). 
We removed  from the catalogue  all the objects which are flagged as belonging to a polluted area  or for which the photometry is possibly contaminated by the light of saturated stars, meaning that only objects in $\cal{A}^{\rm UVISTA}$\&$\cal{A}^{\rm !OPT}$\&$\cal{A}^{\rm COSMOS}$ are kept, according to the notations in Table 7 of \cite{Laigle2016}.  

Photometric redshifts (photo-$z$), stellar masses and absolute magnitudes  have been computed using {\sc LePhare} 
\citep{arnoutsetal2002,ilbert06} with a configuration similar to \citet{ilbert13}.  
The stellar mass completeness of the sample is estimated from the $K_{\rm s}$ magnitude following \cite{pozzettietal2010}. At redshift $z=1.2$, the sample is 90\% complete down to $M_{\rm lim}=10^{9.2} {\rm M}_{\odot}$. However, in this work, only galaxies more massive than $10^{10} {\rm M}_{\odot}$ are used to extract the filament distribution over the redshift range $0.5<z<1.2$.
Because our analysis relies on a global extraction of the filaments -- in contrast to previous works which are based on a local search around groups --, a large enough comoving area is required, which sets the  lower limit of our redshift range to $z\sim 0.5$. The upper limit is defined by the rapid increase of the redshift uncertainties at $z\sim 1.2$ \citep[see Fig.~14 of][]{Laigle2016}\footnote{A consequence of the shift of the Balmer break in NIR broad bands.}.

A large spectroscopic redshift catalogue is also available on the COSMOS field, as the result of the common effort of several spectroscopic follow-up campaigns since 2007 \citep[e.g.][]{lilly07,kartaltepe10,lefevre15,comparat15}. This spectroscopic sample is essential for the calibration of the photometric redshifts and for a better determination of group redshifts. 

 \begin{figure}
\begin{center}
\includegraphics[width=0.45\textwidth]{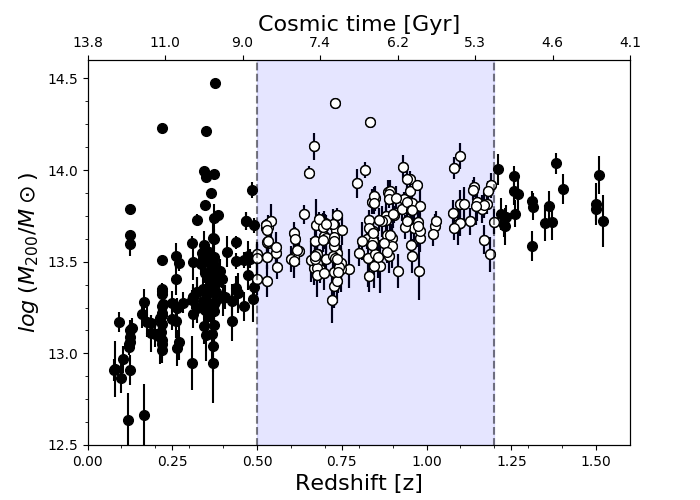}
 \caption{The halo mass of X-ray groups ($M_{200}$) in the COSMOS field as a function of redshift (filled and open circles). The highlighted area represents the groups with a redshift range of $0.5<z<1.2$, which is used in this study (open circles).}
 \label{Fig:mhz}
\end{center}
\end{figure}

\subsubsection{The X-Ray group catalogue}
The initial catalogues of the X-Ray galaxy groups in COSMOS were presented in  \cite{finoguenov07} and \cite{george11}. These catalogues combined the available Chandra and XMM-Newton data (with improvements in the photometric datasets) used for identification of galaxy groups, with confident identification reaching out to $z\sim 1.0$. 
The COSMOS galaxy group catalogue we rely on is a combination of an updated version of the initial group catalogues  and a new catalogue of 73 groups described in \cite{gozaliasl19}  and Gozaliasl et al. in preparation, which combines data of  all X-ray observations from Chandra and XMM-Newton in the 0.5-2 keV band, with robust group identification up to $z\sim 1.53$. However, for the purpose of this study and for the reason explained above, we limit our selection to $z\sim0.5-1.2$ (highlighted area in Fig.~\ref{Fig:mhz}).

Group halo mass is the total mass (commonly called $M_{200}$, but we call it $M_{\rm group}$ in the rest of this paper), determined using the scaling relation $L_{X}-M_{200}$ with weak lensing mass calibration as presented by \citet{leauthaud10}. The radius of the group $R_{200}$ is defined as the radius enclosing $M_{200}$ with a mean overdensity of $\Delta\sim 200$ times the critical background density. Fig.~\ref{Fig:mhz} presents the group mass $\log(M_{200}/{\rm M}_{\odot})$ as function of the redshift and cosmic time.  
\cite{gozaliasl19} discussed the mass completeness of the group sample given the surface brightness limitation of the X-Ray dataset. Over the redshift range $0.5<z<1.2$, the evolution of the group mass limit is weak and lies within the observational uncertainties, being around $\log M_{\rm group}/{\rm M}_{\odot} \sim 13.38$ at $z\sim 0.5$ and  $\log M_{\rm group}/{\rm M}_{\odot} \sim 13.5$ at $z\sim 1.2$. 

The redshift of the group is the redshift of the peak of the galaxy distribution within the group radius, while slicing the lightcone with a redshift step of 0.05. In most cases, this redshift determination is strengthened by the presence of spectroscopic redshifts. The brightest group galaxy (BGG in the following) is identified from the COSMOS2015 photometry as being the most massive galaxy within  $R_{200}$, with a redshift that agrees with that of the hosting group \citep{gozaliasl19}. More  than $\sim80\%$ of the BGGs have secure spectroscopic redshifts. Group centers from the X-Ray emission are determined with an accuracy of $\sim 5\arcsec$, using the smaller scale emission detected by Chandra data. The BGG does not always sit at the peak of the X- Ray centre emission. As described in \cite{gozaliasl19}, the off-central BGG probably reside in groups which are more likely to have experienced a recent halo merger. 

As described in \cite{gozaliasl19}, a quality flag has been assigned to each group depending on the robustness of the extraction and the potential availability of spectroscopic redshift. In our study, we keep only group with a flag of 1, 2, and 3. Over the redshift range $0.5<z<1.2$ and considering only groups with a BGG galaxies more massive than $\log M_{*}/{\rm M}_{\odot}=10$ and in a non-flagged area, we are left with 86 groups containing around 900 galaxies, over 1.38 deg$^{2}$ (or $\sim$62 groups per square degree).

\begin{figure}
\begin{center}
\includegraphics[scale=0.57,trim={2.1cm 0.5cm 0.2cm 0.cm},clip]{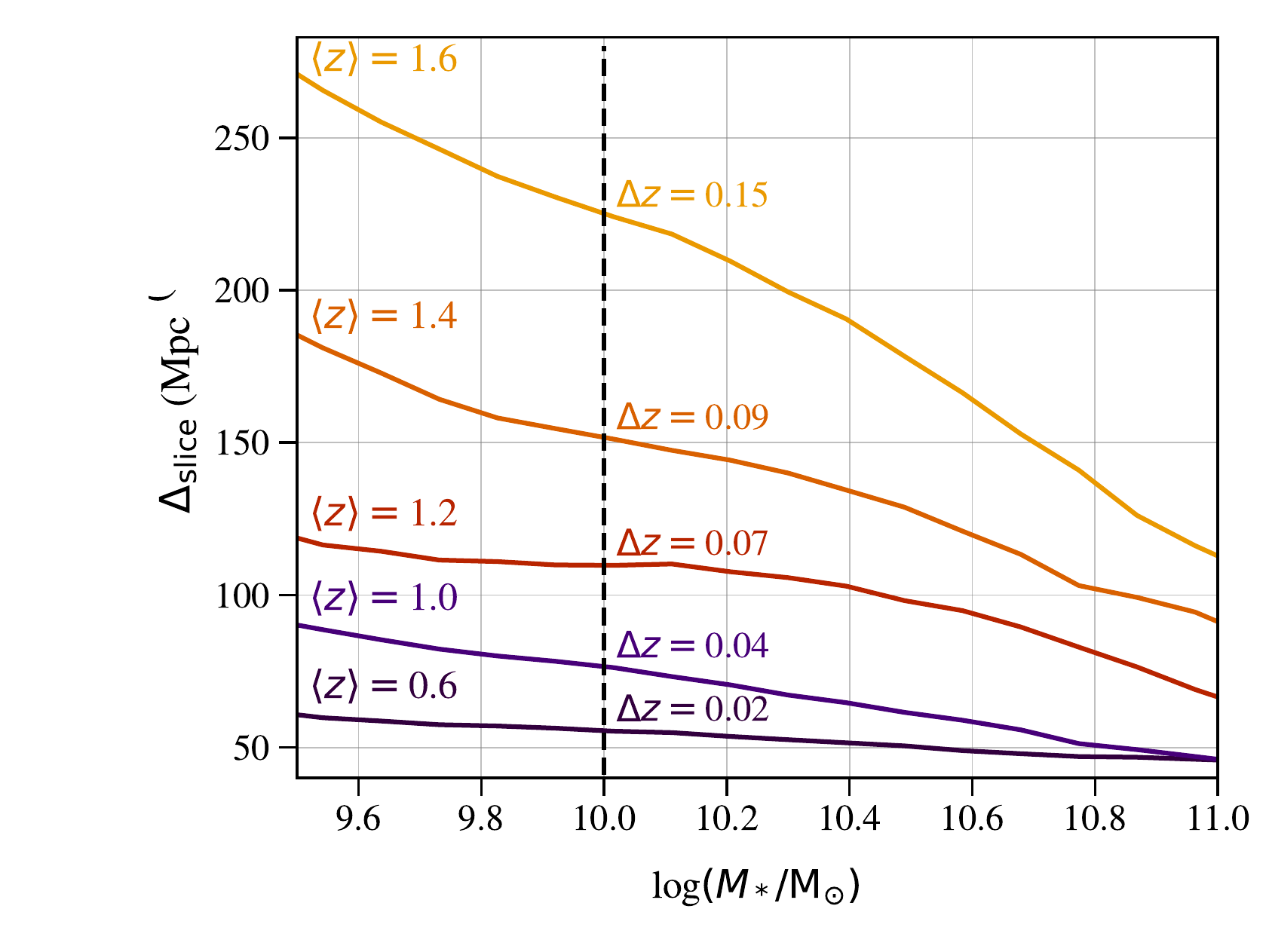}
 \caption{Minimal slice thickness $\Delta_{\rm slice}$ in comoving Mpc which has to be chosen in $\mathcal{C}_{\rm cosmos}$ for the purpose of cosmic web extraction in 2D, as a function of  masses and for different redshifts (from $\langle z \rangle=1.6$  to $\langle z \rangle =0.6$, \textit{yellow} to \textit{dark blue} lines). Here $\Delta_{\rm slice}$ is the comoving thickness  corresponding to   $\Delta{z}=2\times 1\sigma_{z}$, where $\sigma_{z}$ is the redshift uncertainty of the lowest mass galaxies in the sample. Above each line is also indicated $\Delta{z}$ corresponding to the galaxies in the $10^{10} {\rm M}_{\odot}$ bin.}
 \label{Fig:thick}
\end{center}
\end{figure}
\subsection{The {\sc Horizon-AGN} simulation}
\label{subsec:simulation}
The {\sc Horizon-AGN}\footnote{\url{http://www.horizon-simulation.org/}} \citep[][]{dubois14}  cosmological hydrodynamical simulation is used both to assess the quality of the observational measurements and to provide an interpretative framework. Therefore, several catalogues are built from the simulation. 
We first make use of the mock catalogue which has been generated from the  \textsc{Horizon-AGN} simulated lightcone. Two versions of this catalogue are built: $\mathcal{C}_{\rm Hzagn\,2D}^{\rm phot}$ and $\mathcal{C}_{\rm Hzagn\,2D}^{\rm true}$, for which cosmic web filaments are extracted either from the photometric quantities \citep[redshift and masses,][]{laigle19} or intrinsic ones respectively. Comparing the results from these catalogues  allows to quantify the impact of photometric noise and to assess the quality of the connectivity measurement from observations (Section~\ref{Sec:photoz}). \\
Furthermore, in order to interpret the data (Section~\ref{Sec:Discussion}), and in particular to probe how AGN feedback comes into play, an additional catalogue is used:  $\mathcal{C}_{\rm Hzagn\,3D}$, built from the {\sc Horizon-AGN}  snapshot outputs\footnote{The reason why snapshot outputs are used here (instead of the lightcone) is to increase the statistics.}. The simulation is described in the following section, and a summary of all the catalogues  used in this study is presented in Table~\ref{Tab:conf}.
\subsubsection{Description of the simulation}
The  {\sc Horizon-AGN} simulation run has been performed with {\sc ramses}, an adaptative-mesh refinement code introduced by~\citet{teyssier02}.
The size of the simulation box is $L_{\rm  box}=100 \, h^{-1}\,\mathrm{Mpc}$ on a side, and the volume contains $1024^3$ dark matter (DM) particles (which corresponds to a DM mass resolution of $M_\mathrm{DM,res}=8\times 10^7 \, {\rm M}_\odot$). The initially grid is adaptatively refined  down to $1$~physical kpc, leading to a typical number of $6.5\times 10^9$ leaf cells at $z=1$. The refinement is triggered when the number of DM particles becomes greater than 8 or the total baryonic mass reaches 8 times the initial DM mass resolution in a cell.

A uniform UV background is switched on at $z_{\rm  reion} = 10$ following~\cite{haardt&madau96}. 
Gas cools down to $10^4\, \rm K$ via H, He and metals \citep[following][]{sutherland&dopita93}. 
Star particles are created in regions where gas number density is above $n_0=0.1\, \rm H\, cm^{-3}$,  following a Schmidt law: $\dot \rho_*= \epsilon_* {\rho_{\rm g} / t_{\rm  ff}}$,  where $\dot \rho_*$ is the star formation rate mass density, $\rho_{\rm g}$ the gas mass density, $\epsilon_*=0.02$ the constant star formation efficiency  and $t_{\rm  ff}$ the gas local free-fall time.
A subgrid model for feedback from stellar winds and  supernova (both type Ia and  II) is implemented with mass, energy, and metal releases in the surrounding gas.
\textsc{Horizon-AGN} also follows galactic black hole formation, with black hole energy release in either quasar or radio mode depending on the accretion rate  \citep[see][for more details]{dubois12}.
\\
Out of the entire {\sc Horizon} suite, we mostly use the {\sc Horizon-AGN} simulation in this paper.
However, we will briefly compare it to its identical twin without AGN feedback, {\sc Horizon-noAGN} \citep{peirani17}, to highlight the impact of such a feedback on the properties of BGGs.

\begin{figure*}
 \begin{center}
  \includegraphics[height=11.5cm,trim={2.2cm 1.2cm 0cm 0},clip]{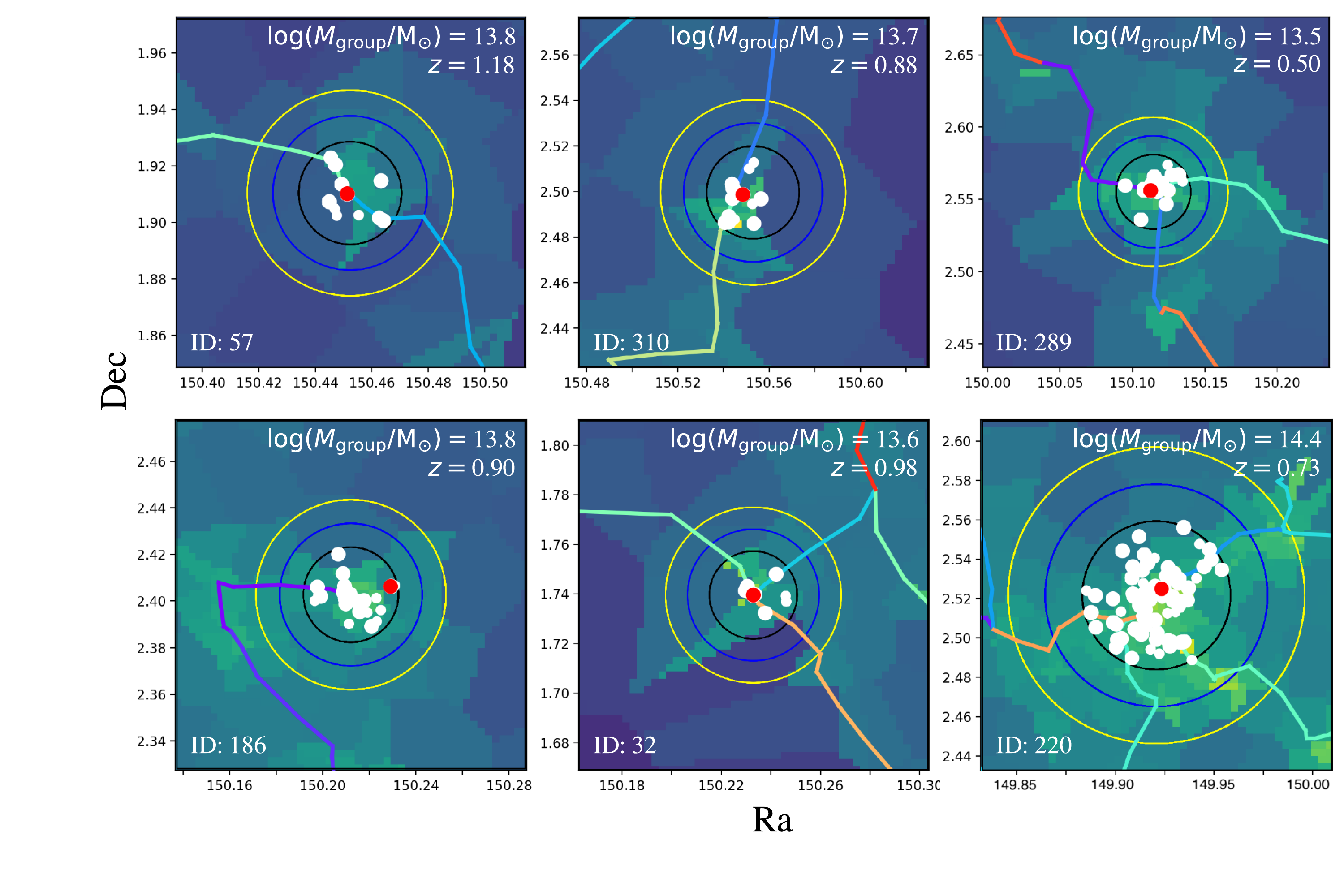}
\end{center}
   \caption{Example of the connectivity of groups of different masses at different redshifts in $\mathcal{C}_{\rm cosmos}$. Each panel is 4 comoving Mpc wide, and the  x- and y-axis indicate respectively right ascension and declination. The black, blue and yellow circles are drawn respectively at $1\times$, $1.5\times$ and $2\times$ the virial radius of the group. The connectivity at a given radius is defined as the number of filaments crossing the corresponding  circle. Galaxies are represented by white disks. Large and small disks correspond to galaxies more massive and less massive than $10^{10} {\rm M}_{\odot}$ respectively. Only galaxies identified as members of the group and with $\log M_{\rm *}/{\rm M}_{\odot} > 9.5 $ are shown. The BGG is in red. Distinct filaments have different colours. The background density is estimated from the Delaunay tessellation.}
   \label{Fig:visualisation}
\end{figure*}
\subsubsection{The snapshot catalogue: $\mathcal{C}_{\rm Hzagn\,3D}$}
In order to identify galaxies from the stellar particles distribution, we run the \textsc{ AdaptaHOP} halo finder \citep{aubert04} on the snapshots. Local stellar particle density is computed from the 20 nearest neighbours, and we keep in the catalogue galaxies with a density threshold  equal to 178 times the average matter density at that redshift.

We then extract DM halos from the DM particle distribution following the same procedure as for galaxies, but with a density threshold of 80 times the average matter density. Only halos with more than 100 particles are kept in the catalogue.  The centre of the halo is temporarily defined as the densest particle in the halo, and then refined with the shrinking sphere method \citep{power03} in order to recursively find the centre of mass of the halo. 

Each galaxy is associated with its closest main halo, and to match the observational definition, the BGG is identified as the most massive galaxy within the virial radius of the main halo.  
In order to increase the statistics, this group catalogue (called $\mathcal{C}_{\rm Hzagn\,3D}$ in the following) is built by joining the data from three snapshot outputs at $z=0.63,\,0.81,\,1.03$ \footnote{which ultimately is a procedure similar to the building of a very low resolution lightcone.}. This results in 1115 groups with $M_{\rm group}>10^{13} {\rm M}_{\odot}$. In order to get the quenching efficiency for the BGG (Section~\ref{Sec:Discussion}), the galaxy catalogue is matched with its counterpart in the {\sc Horizon-noAGN} simulation $\mathcal{C}_{\rm Hznoagn\,3D}$. The matching procedure is fully described in \cite{peirani17} and \cite{beckmann17}. The matching procedure identifies 876 groups of $\mathcal{C}_{\rm Hznoagn\,3D}$ which have a counterpart in $\mathcal{C}_{\rm Hznoagn\,3D}$. For these groups, the quenching efficiency due to AGN feedback is defined for matched galaxies as: $\xi=\log(M_{*\,{\rm Hz-noAGN}}/M_{*\, {\rm Hz-AGN}})$. This quenching efficiency is therefore 0 when the galaxy mass is the same in {\sc Horizon-AGN} and {\sc Horizon-noAGN}.  

{\sc TreeMaker} \citep{tweedetal09} is used to build merger trees from the halo catalogues. Each halo  in a given snapshot at a given redshift is connected to its main progenitors at higher redshift and its child at lower redshift. Merger trees are built over 38 snapshots in the redshift range $z =\left[0.63,5.87\right]$ corresponding to a time step of about $\sim 200$~Myrs. For each group, we look at its merger history by following back in time its merger tree. The halo is assumed to have encountered a major merger of ratio greater than ($1:x$) if the mass ratio of the second to the main progenitors is greater than $1/x$. 

\subsubsection{The lightcone catalogues: $\mathcal{C}_{\rm Hzagn\,2D}^{\rm true}$ and $\mathcal{C}_{\rm Hzagn\,2D}^{\rm phot}$}
\label{Subsubsec:sim}

The {\sc Horizon-AGN} lightcone has been extracted on-the-fly as described in \cite{pichon10}. Gas cells have been replaced by gas particles, and treated  as stars and DM particles. The lightcone projected area is 5 deg$^{2}$ below $z=1$, and 1 deg$^{2}$ above. 
\textsc{ AdaptaHOP} has been run on the lightcone over the redshift range $0<z<4$ with the same method as described above. 
Photometry is  computed for each galaxy in the same filter pass-bands as those available in the COSMOS2015 catalogue, as fully described in \cite{laigle19}. We simply recall here the main features of this catalogue. For a given galaxy, each of its stellar particles is linked to a  single stellar population (SSP) obtained with the stellar population synthesis model of \citet[][]{bruzual&charlot03}, with a Chabrier initial mass function \citep[IMF, ][]{chabrier03}. The galaxy spectrum is the sum of the contributions of the individual SSP. Dust attenuation is also accounted for. 
From this virtual photometry, photometric redshifts and masses have been derived with {\sc Lephare} \citep{arnoutsetal2002} with the same configuration as in COSMOS, and the performance of the simulated catalogue is very similar to the observed one in terms of redshift and mass accuracy \citep{laigle19}. $\mathcal{C}_{\rm Hzagn\,2D}^{\rm true}$ is the catalogue which makes use of the exact galaxy redshifts and masses, while $\mathcal{C}_{\rm Hzagn\,2D}^{\rm photo}$ uses the photometric redshifts and masses. In order to mimic COSMOS data, we keep only groups more massive than $10^{13.3}{\rm M}_{\odot}$ and containing a BGG more massive than $10^{10}{\rm M}_{\odot}$. We are left with 76 groups over 1~deg$^{2}$, which is slightly more than in COSMOS. However cosmic variance is expected to be quite important on these small fields, and the COSMOS group density should be corrected for the small variation of the group mass limit to be fully comparable to the simulated group density.

\subsection{Extraction of filaments and connectivity estimator}
\label{subsec:skeleton}

\begin{figure}
\begin{center}
    \includegraphics[scale=0.59,trim={1.5cm 2.cm 0cm 0.cm},clip]{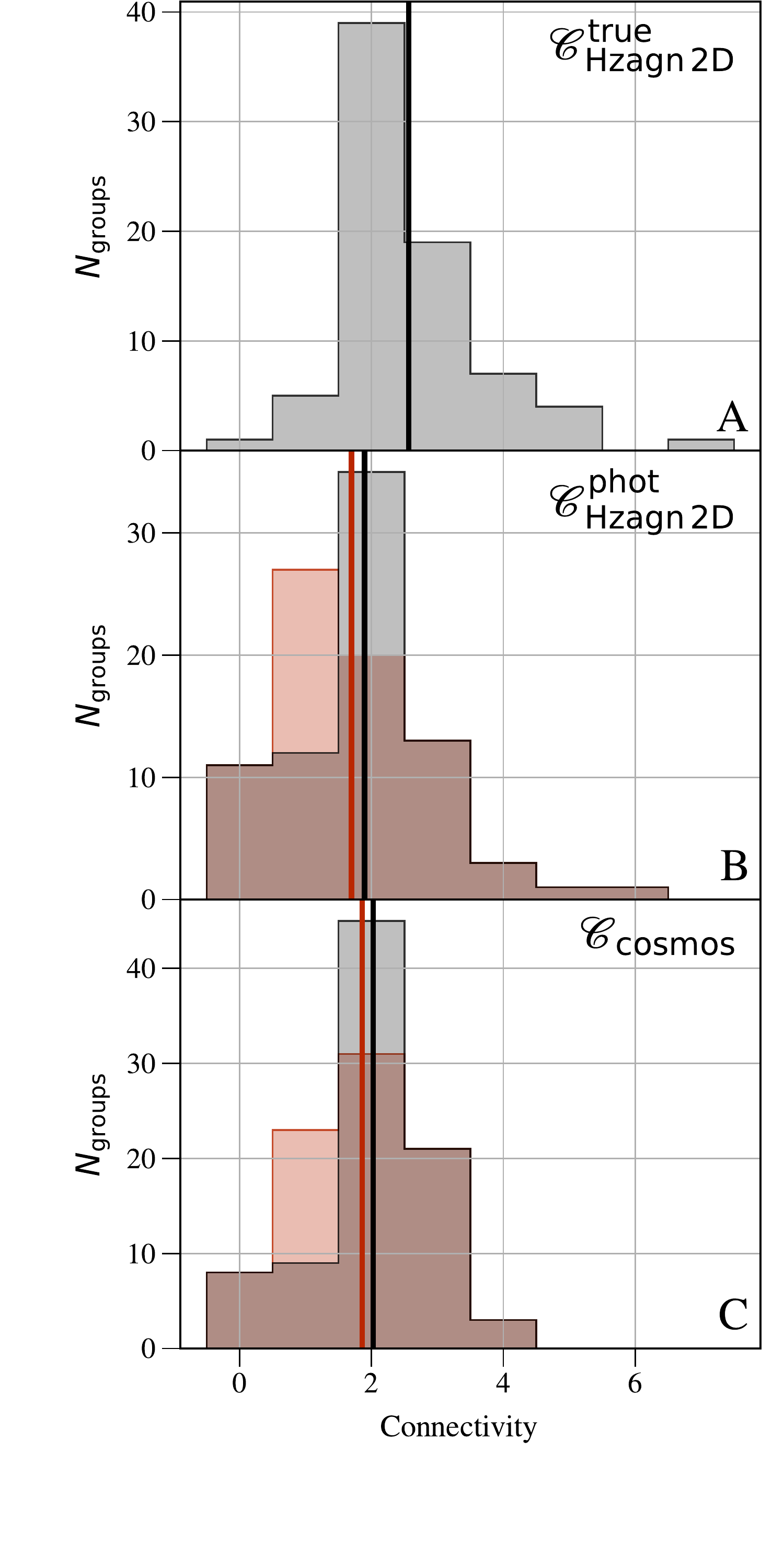}
  \caption{Distribution of group connectivity using a filament extraction based on true-$z$ (\textit{panel A}) and photo-$z$ (\textit{panel B}) from the {\sc Horizon-AGN} simulation. The \textit{panel C} shows the PDF of the connectivity in the COSMOS dataset. The \textit{grey} and \textit{pink} histograms correspond to the distribution with ${\rm 1a}-\pazocal{C}$ groups classed as  2$-\pazocal{C}$ or 1$-\pazocal{C}$ respectively (see text for details). \textit{Black} and \textit{red} lines correspond to the mean of the \textit{grey} and \textit{pink} histograms respectively.}
  \label{Fig:PhotoErr}
  \end{center}
\end{figure}
\subsubsection{Extraction of the filaments in 3D}
To identify the cosmic network from the density field, we use the  persistence based filament tracing algorithm   \citep[{\sc DisPerSE},][]{sousbie111}, which identifies ridges  from the density field  as the special lines connecting topologically robust pairs of saddle-peak critical points. Therefore, the extraction is global in the sense that what determines the presence of a filament at a given location is not only the amplitude of the overdensity with respect to the local background, but also the distribution of  matter on a larger scale (that is, the presence of a saddle point or a peak further away).  In the following, we call ``nodes" the maxima of the density field (where filaments are crossing). 
In \citet{sousbie112}, {\sc DisPerSE} was successfully used to extract filaments around an X-Ray detected group at $z=0.083$. For the 3D extraction of the filaments in {\sc Horizon-AGN}, the density field is reconstructed from the Delaunay tessellation on the entire sample of DM halos \citep{SchappetVandeWeygaert2000}. 

Each filament is defined as a set of connected small segments linking extrema to saddle points. The set of filaments is called the skeleton in the following. Persistence is defined as the difference in density at the two critical points within a pair.
 Expressed in terms of numbers of $\sigma$,  persistence quantifies the significance of the critical pairs  in the Delaunay tessellation of a random discrete Poisson distribution. Thus, the filtering of low-persistence structures ensures that the extraction is  robust with respect to noise. We choose a 5$\sigma$ persistence threshold to extract filaments in 3D from {\sc Horizon-AGN}. 
 
\subsubsection{Extraction of the filaments in 2D}
For the observed catalogue ($\mathcal{C}_{\rm cosmos}$)  and the simulated ones from the lightcone ($\mathcal{C}_{\rm Hzagn 2D}^{\rm true}$ and $\mathcal{C}_{\rm Hzagn 2D}^{\rm phot}$), filaments of the cosmic web are identified from the 2D density field, computed from the Delaunay Tessellation of the galaxy distribution, following \citet{Laigle2018}. However, we introduce two changes compared to this previous extraction.
\paragraph*{Choice of the slice thickness}

We consider slices of a fixed comoving thickness 120~Mpc instead of 75~Mpc (as was chosen in \cite{Laigle2018}). 
The  motivation is to increase the redshift range of the study (up to $z\sim 1.2$ instead of $z\sim 0.9$) while keeping the thickness of the slices calibrated on the typical redshift uncertainty of the lowest mass galaxies in our mass-limited sample ($\Delta_{z}\sim$0.07, that is about 120 cMpc, at $z\sim 1.2$ for $\log M_{*}/{\rm M}_{\odot}\sim 10$). As shown in \cite{Laigle2018}, there is no truly optimal slice thickness for extracting filaments (within some range), allowing for a flexible  choice without degrading the quality of the extraction. It should in practice primarily be guided by the photometric redshift uncertainties of the lowest mass galaxies in the sample. More specifically, the slice thickness is taken as $2\times$ the $\sigma_{z}$ uncertainty\footnote{$\sigma_{z}$ is defined for each galaxy and encompasses 68\% of the PDF($z$) around the median redshift.} of the 5000 lowest mass galaxies in the redshift and mass bin considered. In order to get a constant slice thickness over the whole redshift range, we calibrate the slice thickness on the lowest mass galaxies at $z\sim 1.2$. As shown in Fig.~\ref{Fig:thick}, this implies considering slices of thickness $\sim$120 comoving Mpc. Note that the slice thickness should also be larger than the velocity dispersion in large groups ($\sim 1000$ km/s), which is still small in $\Delta_{z}$ compared to the adopted thickness slices.

The second major change with respect to the analysis in \cite{Laigle2018} concerns the centering of the slices. Because our study is focused on the extraction of filaments around groups, we perform an individual extraction of the filaments for all the groups in our catalogue, each of the groups being put at the center of its own slice. This requirement ensures an optimal extraction of the filaments around all the groups.  
\begin{figure*}
\begin{center}
  \includegraphics[scale=0.58,trim={0.3cm 0.cm 1cm 1.3cm},clip]{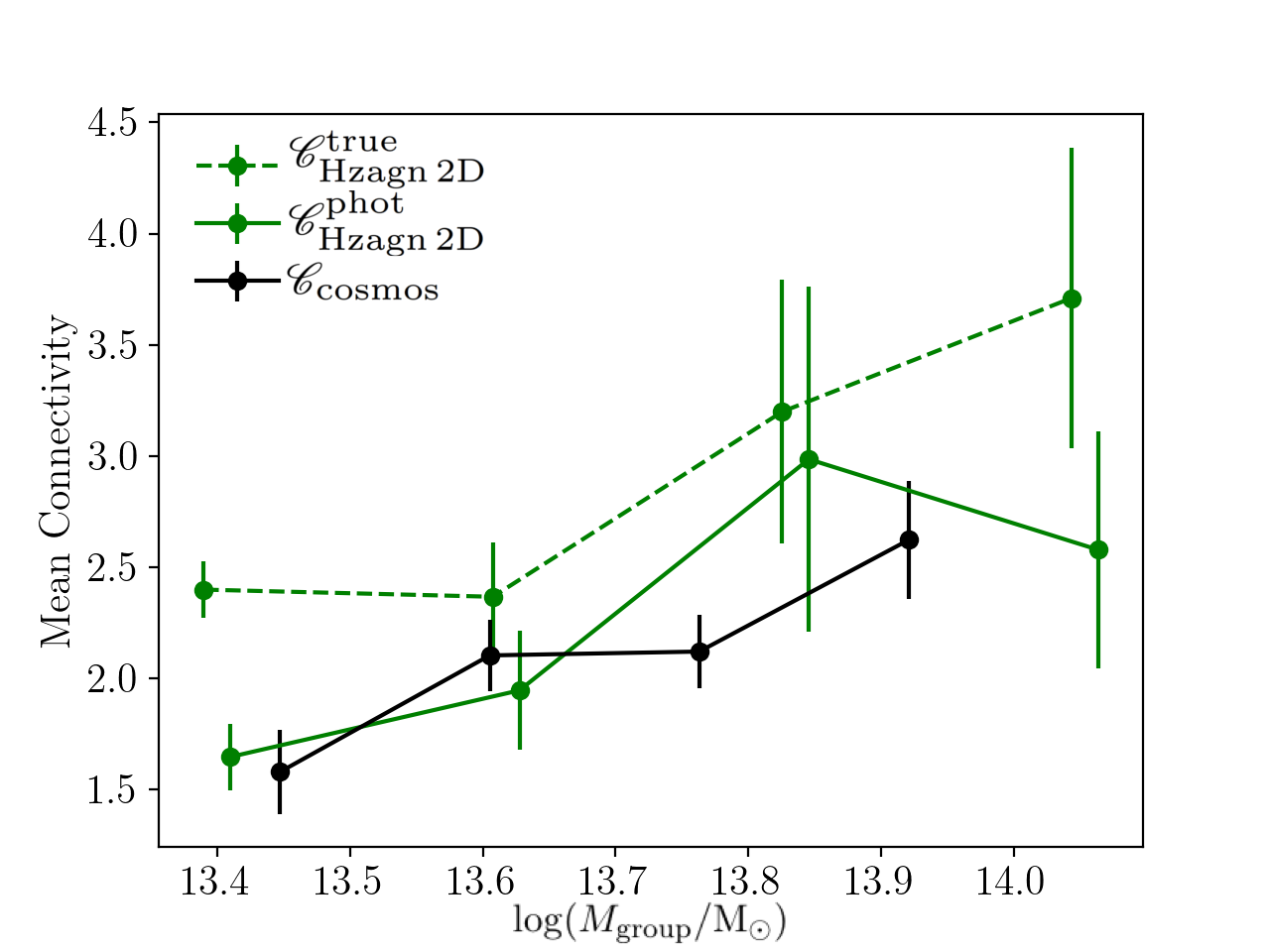}
  \hspace{0.3cm}
   \includegraphics[scale=0.58,trim={0.7cm 0 1cm 1.3cm},clip]{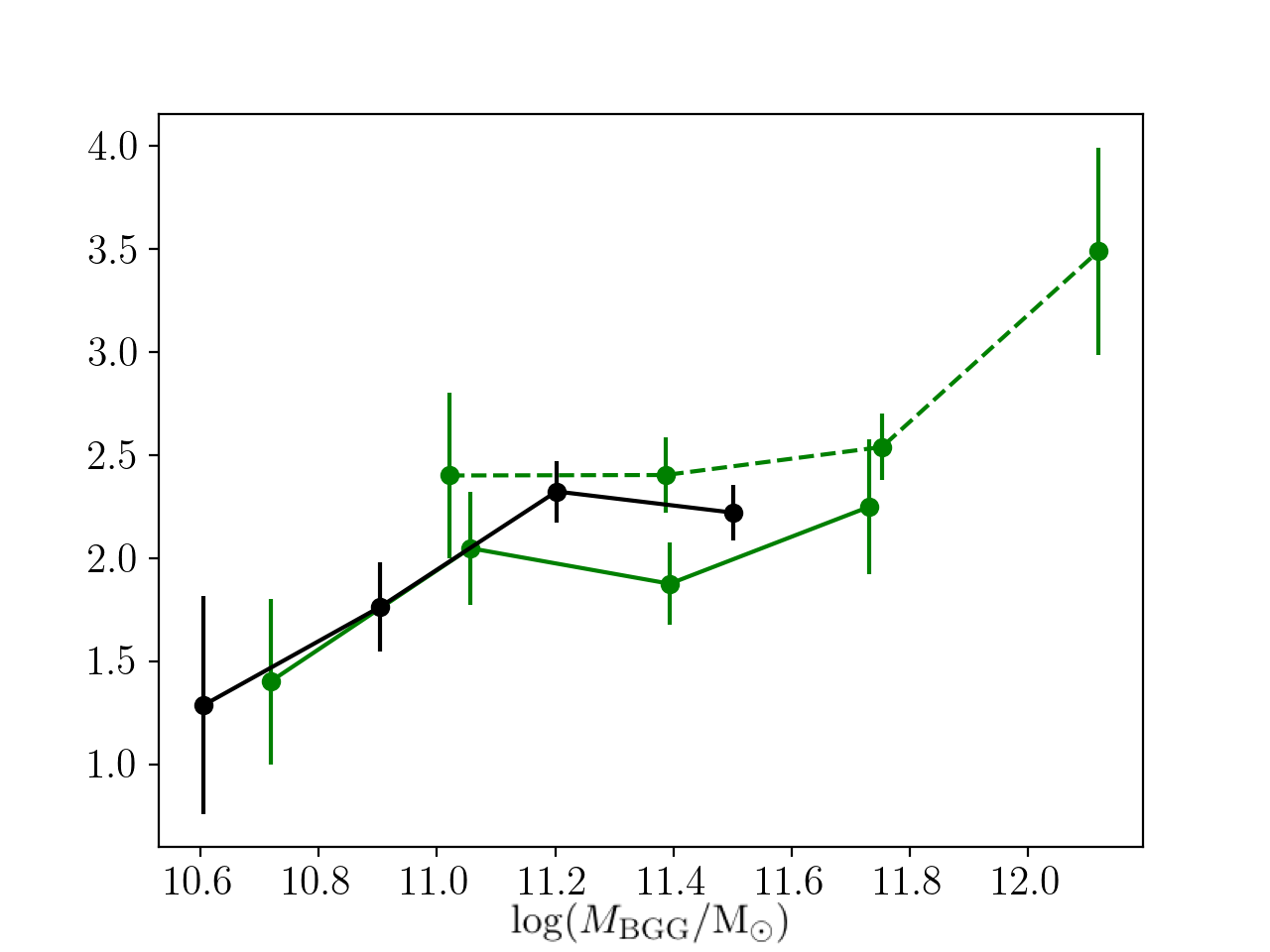}
\end{center}
  \caption{ Mean connectivity plotted against 
  group mass $\protect M_{\rm group}$ (\textit{left}) and BGG mass $\protect M_{\rm BGG}$ (\textit{right})  for COSMOS data (\textit{black} line) and the Horizon-AGN simulation (\textit{green} line) for both photo-z (\textit{solid} line) and true-z (\textit{dashed} line) filament extractions. Photometric masses are also used to compute the relation between mean connectivity and BGG mass in the {\sc Horizon-AGN} mock dataset, which is the reason for the horizontal offset between the \textit{solid} and \textit{dashed} curves on the \textit{right} panel. Errorbars are the error on the mean computed from bootstrap resampling.}
  \centering
  \label{Fig:MassConnec}
\end{figure*}
\paragraph*{Choice of the persistence threshold in 2D}

\cite{Laigle2018} chose a persistence threshold of 2$\sigma$ for slice thickness of 75 Mpc at the same mass complenetess as ours. This choice was justified by comparing in mocks the 2D photometric skeleton with the projected 3D one. As shown on their Fig.~A.1, minimizing the number of unmatched filaments in the reference and in the reconstructed skeletons implies decreasing the persistence threshold with increasing slice thickness (at a given mass completeness). Therefore, we use in this work a persistence threshold of 1.5$\sigma$ \footnote{In addition, a persistence of 1.5$\sigma$ allows to optimally recover the filaments in the mocks ($\mathcal{C}_{\rm Hzagn 2D}^{\rm phot}$) with respect to the intrinsic distribution ($\mathcal{C}_{\rm Hzagn 2D}^{\rm true}$). A 2$\sigma$ persistence however underestimates the connectivity in the mocks $\mathcal{C}_{\rm Hzagn 2D}^{\rm phot}$, especially for highly connected groups. A 1.5$\sigma$ persistence allows to mitigate this underestimation.}. 

\subsubsection{Method to measure the connectivity}
In the skeleton produced by {\sc DisPerSE}, two filaments joining the same node can become extremely close to one another, but still counted separately, as they  both are topologically robust. However, physically they represent a single filament. Therefore, in order to avoid double counting filaments, they are merged in the final skeleton and a bifurcation point is added at the merging location\footnote{In this sense, we measure multiplicity and not connectivity, in the wording of \cite{codis18}.}. \\
The same ID is then attributed to all segments belonging to the same filaments. 
Connectivity $\pazocal{C}$ is subsequently defined as the number of segments with different IDs crossing the 1.5$\times$ virial radius circle around the group center. The same definition is taken in 3D and 2D. In appendix~\ref{sec:counts} we confirm on the observational dataset that varying this radius  does not significantly change the measurement of connectivity and the signal presented hereafter. 

Fig.~\ref{Fig:visualisation} presents six groups in  the COSMOS field, with varying masses, redshifts and connectivity, plotted over the underlying density field estimated from the Delaunay Tessellation. Distinct filaments are shown in different colours.
The BGG is coloured in red, while the other galaxies in the group are coloured in white. On all these panels, the groups are associated with a peak of the galaxy density distribution, i.e. they are sitting on a node of the cosmic web. The  \textit{bottom right} panel is an example of the highest connectivity found (${4}-\pazocal{C}$, i.e. the group is connected to 4 filaments), while the bottom left panel shows an example of the lowest connectivity found that is still associated with a peak in the distribution (${1}-\pazocal{C}$, i.e. the group is connected to only 1 filament). While a $1-\pazocal{C}$ connectivity is unlikely to happen in theory, this case is found in observations because low persistence filaments were removed from the skeleton. Therefore this group is connected to more filaments, however all but one were too noisy to pass the persistence threshold selection.

\subsection{Impact of photo-$z$ errors: insights from mocks}
\label{Sec:photoz}
The degradation of the global filament extraction when working with photo-$z$ has already been investigated in \cite{Laigle2018}. However we focus here on the degradation of the 2D-connectivity measurement, thanks to the comparison of $\mathcal{C}_{\rm Hzagn 2D}^{\rm phot}$ and $\mathcal{C}_{\rm Hzagn 2D}^{\rm true}$.  The grey histograms on the top and middle panels of Fig.~\ref{Fig:PhotoErr} present the PDF of group connectivity from the {\sc Horizon-AGN} simulation, either using the true galaxy redshift (\textit{panel A}, $\mathcal{C}_{\rm Hzagn 2D}^{\rm true}$) or the photo-$z$ (\textit{panel B}, $\mathcal{C}_{\rm Hzagn 2D}^{\rm phot}$) to extract the 2D cosmic web. The vertical solid black lines give the mean of each distribution.
Photo-$z$ errors tend to decrease the global connectivity, as they introduce shot noise in the density measurement and therefore disconnect the field.

In practice, when the measure is performed from photo-$z$ ($\mathcal{C}_{\rm cosmos}$ and $\mathcal{C}_{\rm Hzagn 2D}^{\rm phot}$), a non-negligible number of groups are found which are not connected to filaments. Another non-negligible number of groups are embedded in filaments but not associated with a peak of the galaxy density distribution above the persistence threshold (1a$-\pazocal{C}$). Finally, others are associated with a peak of the galaxy distribution, but associated to only 1 filament (1b$-\pazocal{C}$), because the other filaments have been removed when filtering the pairs of critical points below the persistence threshold. These cases are unlikely to occur when connectivity is computed from true galaxy redshifts and are mostly due to noise in the photo-$z$. Indeed, their fractions is almost null in $\mathcal{C}_{\rm Hzagn 2D}^{\rm true}$ (\textit{top} panel), and in particular no 1a$-\pazocal{C}$ case is found.\footnote{To confirm that these situations are created by photo-$z$ uncertainties and are not driven by some physical properties of these groups, we compare the distributions of group mass and BGG mass for the low connectivity sample (${0}-\pazocal{C}$, ${\rm 1a}-\pazocal{C}$ and ${\rm 1b}-\pazocal{C}$ ) with the ones for the high connectivity sample ($1<\pazocal{C}$). We do not observe any significant bimodality between the distributions of both samples, the low-$\pazocal{C}$ sample behaving like the low-mass tail of the high-$\pazocal{C}$ one.}
\\
It  is debatable  if groups embedded in filaments but not associated with a peak of the density distribution (1a$-\pazocal{C}$) should be counted as 1$-\pazocal{C}$ or 2$-\pazocal{C}$, as there are formally 2 filaments branching out of these groups. The comparative analysis from the detailed  mocks suggests these groups should sit at a density peak in the absence of photo-$z$ errors.  Therefore in the following we make the choice to consider them as 2$-\pazocal{C}$ (\textit{grey} histograms on Fig.~\ref{Fig:PhotoErr}) instead of 1$-\pazocal{C}$ (\textit{pink} histograms).  Appendix~\ref{App:v1v2} discusses this issue in more details. For each measurement based on $\mathcal{C}_{\rm cosmos}$ presented in the following, we also checked that the result is not strongly dependent on this choice.

\begin{figure}
  \includegraphics[scale=0.59,trim={1.3cm 0.cm 0.5cm 0.cm},clip]{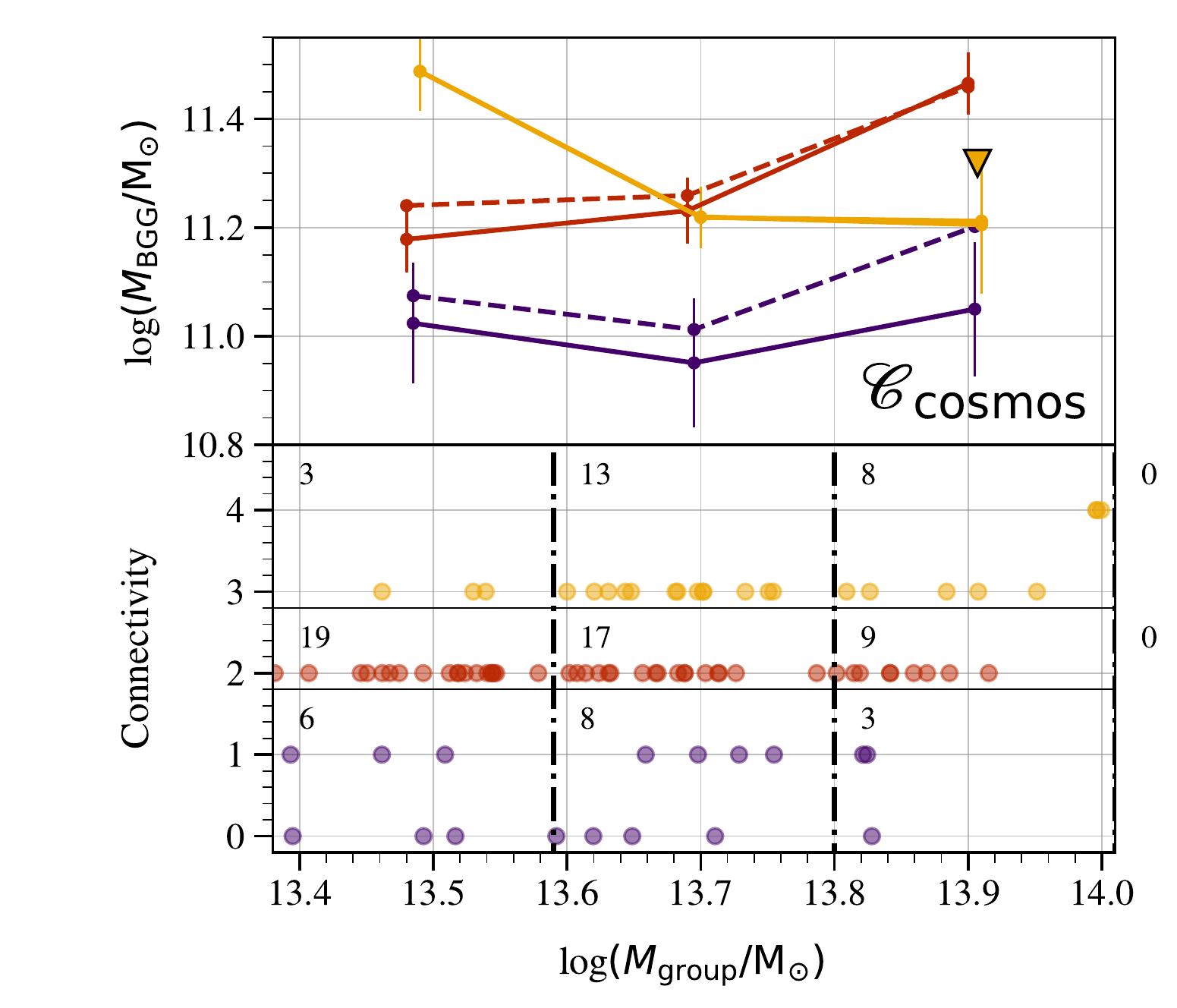}
  \caption{ The BGG mass  against group mass for groups with connectivity smaller than 2 (\textit{blue} line), connectivity of 2 (\textit{red} line) and connectivity greater than 2 (\textit{yellow} line). Two groups of connectivity greater than 2 have their mass higher than 10$^{14}{\rm M}_{\odot}$. When they are included in the measurement of the last mass bin, we get the \textit{yellow triangle} marker. The dashed lines correspond to the same measurements, but when classifying the 1a$-\pazocal{C}$ as 1$-\pazocal{C}$ instead of 2$-\pazocal{C}$.
  The \textit{bottom} panel shows the distribution of groups in bins of connectivity and group mass. In the corner of each 2D-bin is indicated the number of groups in this bin.}
  \centering
  \label{Fig:centralratio}
\end{figure}

\section{Group connectivity in COSMOS} 
\label{Sec:Results}
%
\begin{figure*}
\begin{center}
  \includegraphics[scale=0.58,trim={2cm 0.2cm 0.5cm 0cm},clip]{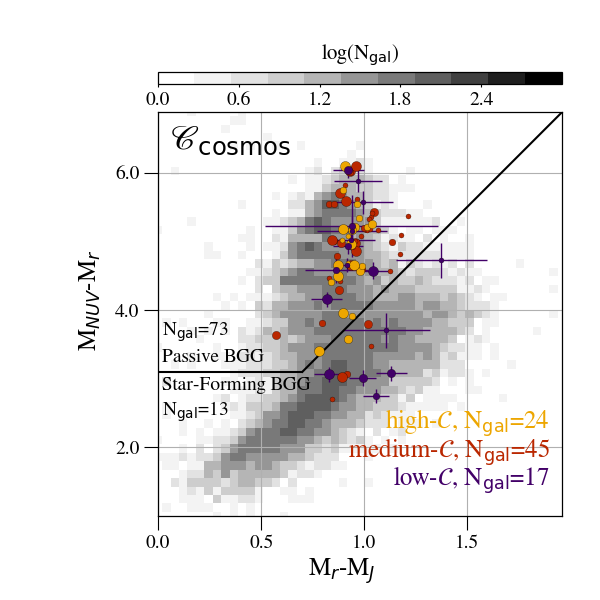}
\hspace{0.3cm}
  \includegraphics[scale=0.6,trim={2.4cm 0.5cm 0.2cm 0.cm},clip]{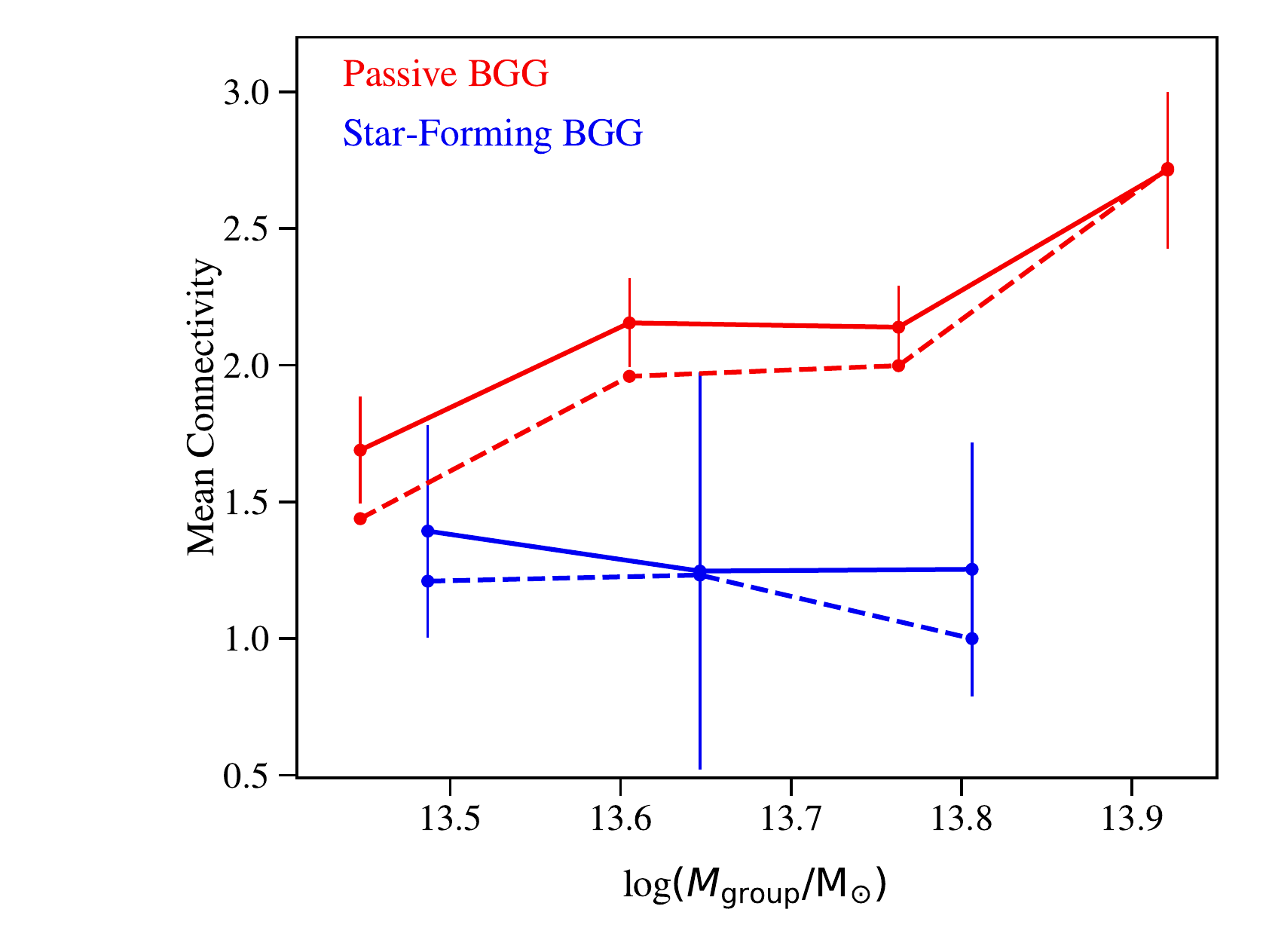}
    \end{center}
  \caption{ \textit{Left}: $M_{NUV}-M_{r}$/$M_{r}-M_{J}$ distribution for the entire galaxy population in 0.5$<z<$1.2 and  $M_{*}>10^{10}{\rm M}_{\odot}$ in COSMOS (\textit{grey} area), and for the BGGs only in the same redshift and mass ranges (coloured circles). Circles are coloured according to their connectivity (from low to high connectivity, \textit{blue} to \textit{yellow} respectively, according to the binning shown in Fig.~\ref{Fig:centralratio}), and their sizes reflect the group masses. The solid line is the usual separation between passive and star-forming galaxies (see text). For clarity, only errorbars for the low-connectivity BGGs are shown. \textit{Right}:
  Mean connectivity as a function of group mass in COSMOS, for groups with a passive BGG (\textit{red} line) and groups with a star-forming one (\textit{blue} line). The dashed lines correspond to the same measurements, but when classifying the 1a$-\pazocal{C}$ groups as 1$-\pazocal{C}$ instead of 2$-\pazocal{C}$.
  }
  \centering
  \label{Fig:SHMR}
\end{figure*}

The \textit{grey} histograms on the \textit{middle} and \textit{bottom} panels of Fig.~\ref{Fig:PhotoErr} show the distribution of the connectivity in the full redshift range $0.5<z<1.2$ in the {\sc Horizon-AGN} and COSMOS datasets. We measure $\langle \pazocal{C} \rangle=2.02$~; ${\rm RMS}\left(\pazocal{C}\right)=0.92$ in $\mathcal{C}_{\rm cosmos}$, $\langle\pazocal{C}\rangle=2.56$~; ${\rm RMS}\left(\pazocal{C}\right)=1.10$ in $\mathcal{C}_{\rm Hzagn 2D}^{\rm true}$ and $\langle\pazocal{C}\rangle=1.89$~; ${\rm RMS}\left(\pazocal{C}\right)=1.18$ in $\mathcal{C}_{\rm Hzagn 2D}^{\rm phot}$.

\subsection{Mean connectivity and group mass dependency}

The mean group connectivity as a function of group mass can now be measured. In both COSMOS and the {\sc Horizon-AGN} datasets, the group mass $M_{\rm group}$ is defined as the total mass, i.e. the sum of DM and baryonic mass. The \textit{left} panel of Fig.~\ref{Fig:MassConnec}  displays this measurement in the mock ($\mathcal{C}_{\rm Hzagn 2D}^{\rm phot}$ and $\mathcal{C}_{\rm Hzagn 2D}^{\rm true}$, \textit{green} solid and dashed lines resp.) and in COSMOS ($\mathcal{C}_{\rm cosmos}$, \textit{black} line). Group mass bins are split to contain an approximately equivalent number of groups in each bin. 
Unsurprisingly and as discussed above, the measurement performed with true-$z$ lies above the one with photo-$z$. The COSMOS and photo-$z$ {\sc Horizon-AGN} datasets are however in very good agreement. 
As expected from theoretical predictions \citep{codis18}, more massive groups have, on average, a higher connectivity.

\subsection{Impact of connectivity on BGG mass assembly}
The impact of the connectivity on the mass assembly of the BGG is now investigated. Our purpose is to quantify if there is any correlation between connectivity and the BGG properties (mass and type)  beyond the trend driven by the group mass (which scales with connectivity). The adopted strategy is therefore to look at each of these quantities in bins of group mass. 

\subsubsection{Mass of the BGG}
The overall evolution of mean connectivity as a function of BGG mass is first measured. This result is displayed in the \textit{right} panel of Fig.~\ref{Fig:MassConnec}. As a consequence of  the assumptions made at the SED-fitting stage when computing masses from photometry \citep{laigle19}, the BGG photometric masses are systematically underestimated with respect to their intrinsic mass in {\sc Horizon-AGN}, which is the reason for the horizontal offset between the \textit{solid} and \textit{dashed green} curves. Taking into account this systematic, as well as the lowering of the connectivity when computed from photometric redshifts, brings  $\mathcal{C}_{\rm Hzagn 2D}^{\rm phot}$ in good agreement with $\mathcal{C}_{\rm cosmos}$ within the errorbars (\textit{green} and \textit{black} curves respectively). As expected, the connectivity is higher when the BGG is more massive. At first order this result is a natural consequence of the scaling of group mass with connectivity, given the known  correlation between group mass and  mass of the BGG.

In order to investigate if  there is an additional correlation between connectivity and BGG mass (beyond the effect driven by group mass), 
the \textit{top} panel of Fig.~\ref{Fig:centralratio} presents the BGG mass as a function of group mass in 3 different connectivity bins. The \textit{bottom} panel presents how groups are distributed within the mass and connectivity bins. Group mass bin are uniformly distributed in logarithm scale between $10^{13.38} {\rm M}_{\odot}$ and $10^{14.02} {\rm M}_{\odot}$. 
 At fixed group mass in the low mass bins ($M_{\rm group}\lesssim 10^{13.7}{\rm M}_{\odot}$), there is a significant trend for the BGG to have higher mass at higher connectivity. 
In the highest mass bins, the mass of the BGG for groups with $\pazocal{C}\geq 3$ (\textit{yellow solid} line) is on average lower as those with $\pazocal{C}=2$ (\textit{red} line). Note that two groups in the highest connectivity bin have their mass higher than $10^{14} {\rm M}_{\odot}$, and are therefore not included by default on the highest mass bin. Including them (\textit{yellow triangle} marker) in the highest mass bin sensibly biases the mass distribution of groups in this  connectivity bin (compared to the medium connectivity bin). In despite of this, the mean BGG mass is still lower than the one in the medium connectivity bin (compare the last \textit{red circle} marker and the \textit{yellow triangle} marker).
We also checked that this measurement is not strongly dependent on the choice we made to consider 1a$-\pazocal{C}$ groups as 2$-\pazocal{C}$ (compare \textit{solid} and \textit{dashed} lines). 

Although this result needs more statistics to be confirmed, it suggests that low-mass groups build more efficiently the mass of their BGG at high connectivity, while in high mass group there is a stagnation of the BGG mass assembly at high connectivity.  Altogether, this suggests that connectivity might trace different assembly histories of groups.

\subsubsection{Type of the BGG}
To further probe the link between  connectivity and the mass assembly of the BGG, 
the relationship between connectivity and group mass is investigated in COSMOS by splitting the group population as a function of the type of their BGG. The classification between star-forming and passive galaxies is performed according to the $M_{NUV}-M_{r}$/$M_{r}-M_{J}$ diagram as shown in the {\textit{left} panel} of Fig.~\ref{Fig:SHMR} \citep[as done in][]{Williams2009,ilbert13,Laigle2016}. Quiescent galaxies are those with $\left[M_{NUV}-M_{r}>3(M_{r}-M_{J})+1\right] \& \left[M_{NUV}-M_{r}>3.1\right]$.
Most of the BGG are passive, however 13 of them are classified as star-forming, with group mass ranging from $10^{13.41}{\rm M}_{\odot}$ to $10^{13.89}{\rm M}_{\odot}$. In this mass range, the mean connectivity of groups with a passive BGG is $\langle\pazocal{C}_{\rm pass}\rangle=$2.09, while it is $\langle\pazocal{C}_{\rm sf}\rangle=$1.3  for the groups with a star-forming BGG. 
In the {\textit{left} panel} of Fig.~\ref{Fig:SHMR}, circles representing BGG are color-coded according to the connectivity of their group, and their size scales with the mass of the group. Although some groups with a passive BGG have a low connectivity, no star-forming BGG except one (in the vicinity of the separation line) are in the highest connectivity bin.

As a complement, the \textit{right} panel of Fig.~\ref{Fig:SHMR} presents the mean connectivity, as a function of group mass, of groups with either a passive or a star-forming BGG. At a fixed group mass, the mean connectivity of groups with a passive BGG is higher than for groups with a star-forming BGG. 
This result suggests that higher connectivity implies a higher chance for the BGG to be quenched. 
The analysis of the {\sc Horizon-AGN} simulation in the following Section~\ref{Sec:Discussion} will help interpreting this result.

\section{Interpretation and discussion}
\label{Sec:Discussion}
\subsection{Summary of the observational findings}
The observational results presented in the previous section confirm first that connectivity scales as a function of group mass. 
The agreement of this result with theoretical predictions is also indirectly a quality assessment of the filament extraction from photo-$z$ redshift in COSMOS. 
The relation between BGG mass and connectivity is at first order a natural consequence of the relation between group mass and connectivity. 

We also investigated the correlation between  connectivity and the properties of the BGG (mass and type) {beyond the trend which could be deduced from the correlation between connectivity and group mass}.  
We found that, in low mass groups ($M_{\rm group} \lesssim 10^{13.7} {\rm M}_{\odot}$), the mass of the BGG is on average higher if the connectivity is higher. This trend reverses above this mass,  with BGGs in  highly connected groups ($\pazocal{C}\geq 3$) being on average less massive than their counterparts in groups of the same mass  (Fig.~\ref{Fig:centralratio}). In addition, even if passive galaxies are found both at low and high connectivity, BGGs hosted by highly connected groups are always passive (Fig.~\ref{Fig:SHMR}), suggesting a quenching process connected with (but not necessarily caused by) a high connectivity. 
   
\begin{figure}
\begin{center}
\includegraphics[scale=0.59,trim={1.5cm 2.5cm 0cm 0.cm},clip]{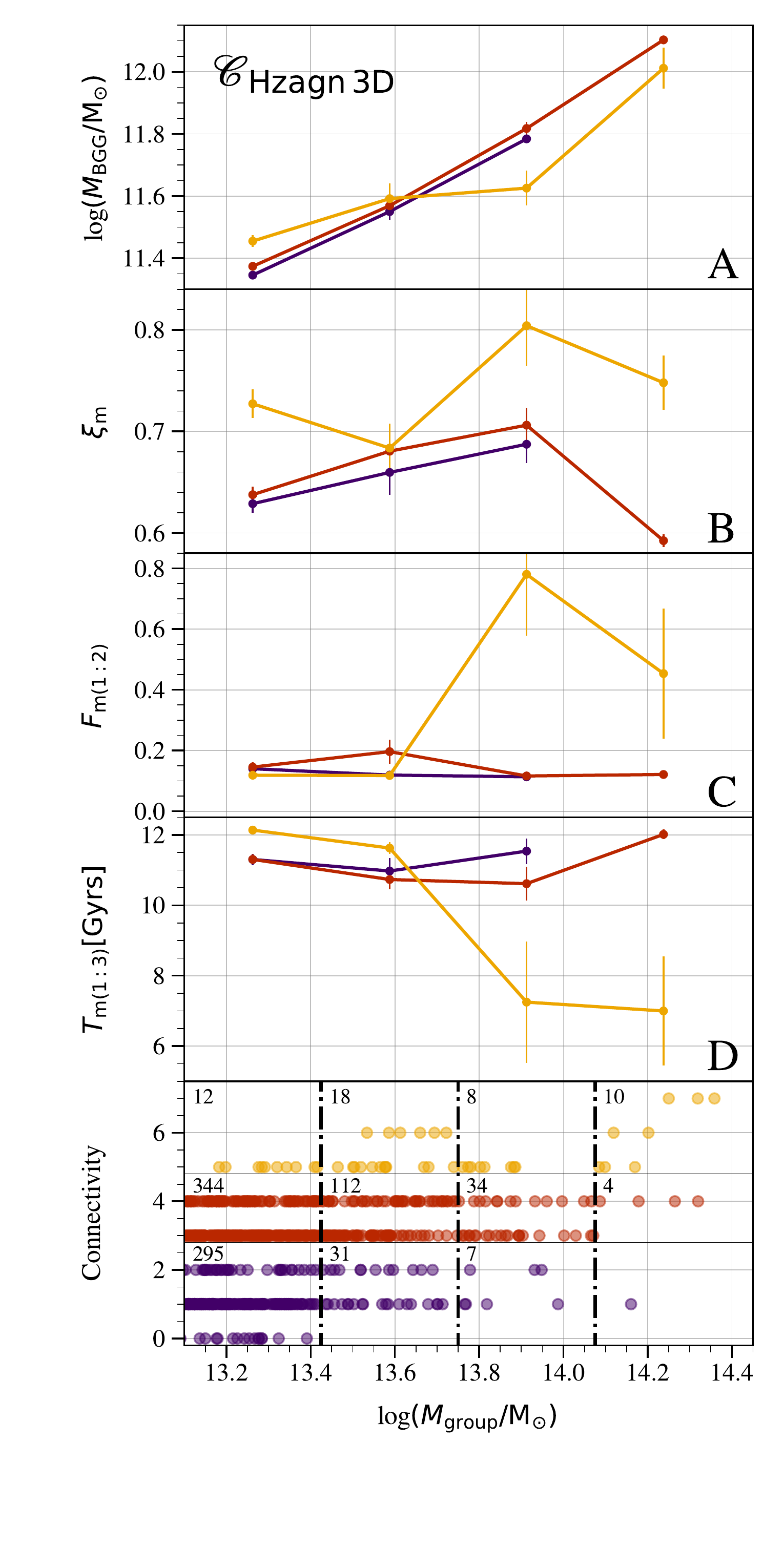}
\end{center}
  \caption{Group properties as a function of group mass in {\sc Horizon-AGN}, in 3 different connectivity bins. \textit{Panel A}: mean BGG mass; \textit{Panel B}: mean quenching efficiency due to AGN feedback; \textit{Panel C}: fraction of groups with a major merger of ratio (1:2) during the last 6 Gyrs; \textit{Panel D}: mean time in Gyrs since the last major merger of ratio (1:3). \textit{Bottom panel}: the distribution of groups in bins of connectivity and group mass. The number of groups in each bin is indicated on the top left of the cell. The groups of a given colour set are used to build the curves of the same colour in the panels $\protect A$, $\protect B$, $\protect C$ and $\protect D$. }
 \centering
 \label{Fig:all}
\end{figure}

Several physical processes  might be responsible for this trend. 
First, assuming  gas accretion rate scales with connectivity, one can expect that  higher gas infall along more filaments triggers stronger feedback within the BGG. This feedback could in turn  prevent stellar mass growth. This interpretation is appealing, as it could both explain the trend at low group mass (higher connectivity groups host more massive galaxies, because more matter is efficiently accreted onto the BGG) and the reverse of this trend at high group mass (the higher accretion rate in high connectivity triggers stronger AGN feedback).
These findings might also be an outcome of assembly bias, in which 
higher connectivity might be a tracer of recent group mergers. As a matter of fact, when two groups of connectivity $n$ and $m$ respectively are merging along a filament, their others filaments are likely to not merge in the same timescale. The net connectivity resulting of the merger will therefore be $n+m-2$. As long as both $n$ and $m$ are greater than 2, the resulting connectivity will be higher than the mean connectivity of the 2 progenitors. Mergers are violent events. The merger of two halos -and consequently their two central galaxies- might shock-heat the gas, but also increase the velocity dispersion, enhance black hole growth and in particular be the reason for stronger AGN feedback \citep[e.g.][]{DiMatteo05}. Although mergers can temporally trigger bursts of star formation \citep[e.g.][]{DiMatteo07,martin17},  the merger-enhanced AGN activity can in the long term durably quench the galaxies \citep[e.g.][]{Springel05,dubois16}. 
This interpretation from mergers is consistent with previous works finding that SFR is enhanced in isolated groups compared to groups embedded  in larger structures (e.g. \citealp{scudder12}, where AGN however have been explicitly excluded from the sample; \citealp{luparello15}).

\begin{figure}
\includegraphics[scale=0.6,trim={2.4cm 0.5cm 0.2cm 0.cm},clip]{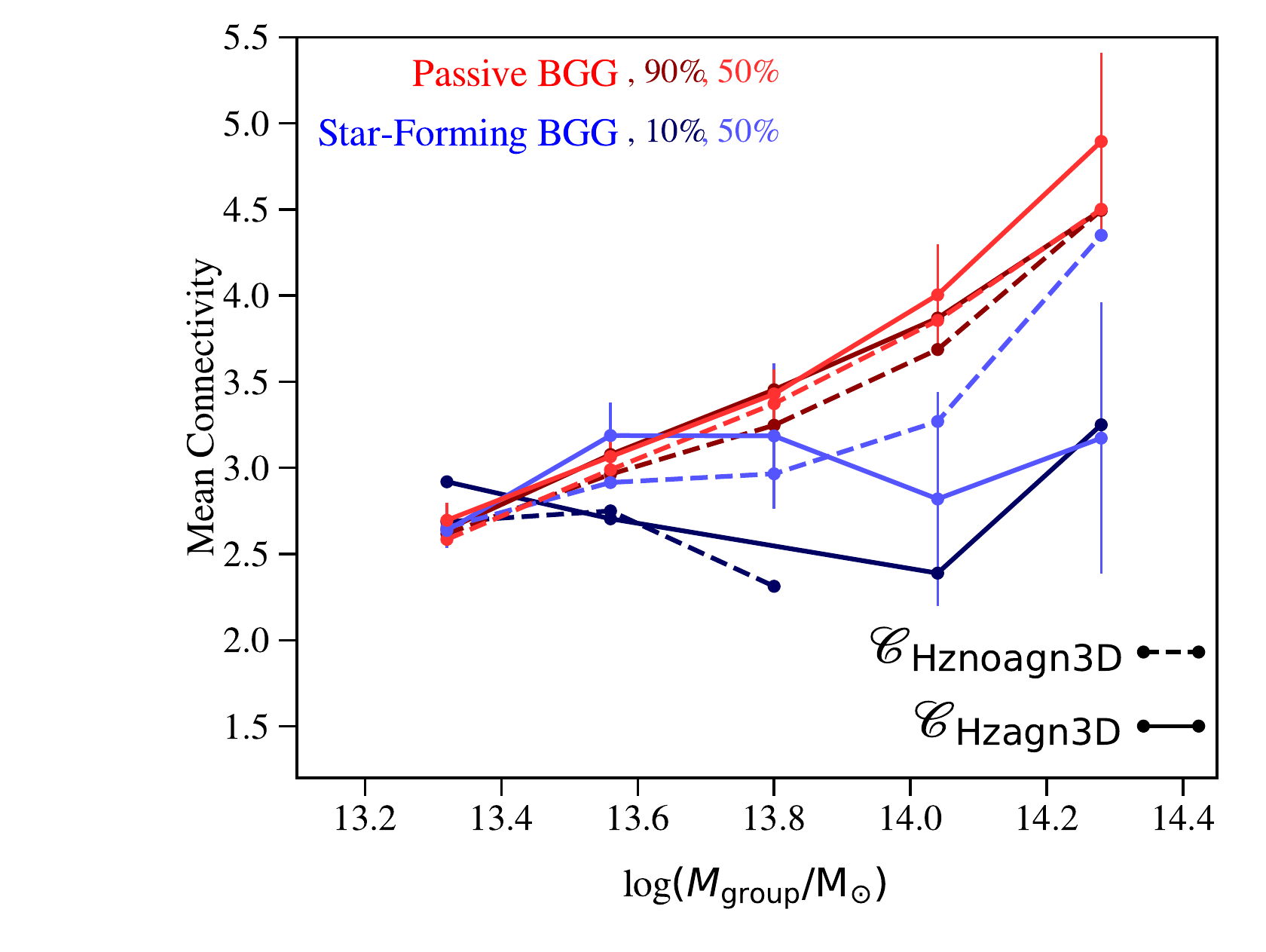}
  \caption{Mean connectivity as a function of group mass in the {\sc Horizon-AGN}  (\textit{solid} line) and {\sc Horizon-noAGN} (\textit{dashed} line) for passive (\textit{red}) and star-forming (\textit{blue}) BGGs, while chosen the two extreme thresholds (from dark to light colours) to separate the passive population from the star-forming one (see text for details). For clarity, only errobars on the lighest curves are shown.}
 \centering
 \label{Fig:sSFRsim}
\end{figure}
\subsection{Interpretation from {\sc Horizon-AGN}}
We also conduct complementary measurements in the {\sc Horizon-AGN} simulation with the $\mathcal{C}_{\rm Hzagn 3D}$ catalogue in order to understand the observational results.

Mean connectivity is first measured as a function of group mass both for groups with a passive and star-forming BGG. Separating passive and star-forming BGGs from the $M_{NUV}-M_{r}$/$M_{r}-M_{J}$ diagram is less easy in {\sc Horizon-AGN} than in COSMOS, as there is residual star-formation even in the passive systems \citep{kaviraj16}. For this reason we decided to sort galaxies based on their specific Star-Formation Rate (sSFR) and to define the star-forming sequence as the galaxies with the highest sSFR and the passive sequence as the galaxies with the lowest sSFR (note that the trend does not change when galaxies are sorted according to their SFR rather than their sSFR). Fig.~\ref{Fig:sSFRsim} shows the mean connectivity as a function of group mass in $\mathcal{C}_{\rm Hzagn 3D}$ (\textit{solid} line), for the groups with a passive (\textit{red} lines) and star-forming (\textit{blue} lines) BGG, while choosing two extreme thresholds (in percentage of the total population) to separate the star-forming from the passive BGGs. Overall, whatever the chosen threshold, a trend qualitatively similar to the observations is found in the simulation: the mean connectivity of groups with a star-forming BGG is lower than the one with a passive BGG, especially at high group mass. However, we note that the typical group mass for which a difference between the mean connectivity of the star-forming and passive populations is measurable ($\log M_{\rm group}>10^{13.8}{\rm M}_{\odot}$) is higher than in $\mathcal{C}_{\rm cosmos}$ ($\log M_{\rm group}>10^{13.5}{\rm M}_{\odot}$). We carry out the same measurement in the simulation without AGN feedback (\textit{dashed} line) and a weaker trend emerges, suggesting  that AGN feedback is important to durably quench the galaxies at high connectivity, but might not be the only driver of the trend.

We turn then to measuring the mean BGG mass as a function of group mass, in bins of connectivity (\textit{panel A} of Fig.~\ref{Fig:all}). As found in observations, the high-mass groups in the highest connectivity bins tend to have a lower mean BGG mass than those in the intermediate connectivity bins, although the trend is not very significant.  We do not find a significant reversal of this trend at low group mass (the fact that more connected groups have a more massive BGG at fixed group mass). 
 \\
In order to understand the role AGN feedback plays in driving this trend, the mean quenching efficiency due to AGN feedback is also displayed as a function of group mass, in distinct connectivity bins. A strong signal for quenching efficiency to be higher in high connectivity groups is detected (\textit{panel B} of Fig.~\ref{Fig:all}). 
\\
We then measure, in each connectivity bins, the fraction of groups with a recent major merger of ratio greater than ($1:2$) (\textit{panel C} of Fig.~\ref{Fig:all}). There is a significant signal for this fraction to be higher at higher connectivity. Finally, we measure in bins of group mass the average time spent since the last major merger with a mass ratio greater than ($1:3$) (\textit{panel D} of Fig.~\ref{Fig:all}). At a given group mass, there is a significant signal for the halos in the highest connectivity bin to have encountered a major merger more recently than in the lower connectivity bins.

Consistently with the observational results at high group mass, our measurements support a scenario in which high connectivity is associated with a stagnation of star formation.  
In addition, we found higher AGN quenching efficiency at higher connectivity. Furthermore, the groups with the highest connectivity are found to have more likely encountered a major merger recently. These measurements are consistent with a scenario in which mergers might be responsible for populating the high connectivity bin. As mergers favour black hole growth \citep{hopkins08,schawinski10,dubois16} and AGN activity, AGN feedback might be stronger in high connectivity groups and more efficient in quenching the BGG.  Further measurements, including the black hole growth and AGN activity, gas inflows and outflows, BGG morphology and redshift evolution of the signal, are still required to fully confirm this scenario. These measurements are beyond the scope of this paper and will be carried out in a upcoming work.

\section{Summary and conclusion}
\label{Sec:Conclusion}
\vskip 0.1cm

The number of filaments connected to a given group (within 1.5 times the virial radius) has been  measured in COSMOS around X-Ray detected groups  in the redshift range $0.5<z<1.2$. Its correlation with group mass, BGG mass and type has been investigated.
The summary of our findings is as follows:
\begin{itemize}
\item[(1)] In COSMOS, groups are found to be connected on average to 2 filaments, and up to 4 filaments. A  small fraction of X-ray-detected groups do not lie at a peak of the galaxy distribution as extracted by {\sc DisPersE};
\item[(2)] A mock photometric catalogue extracted from the {\sc Horizon-AGN} lightcone has been used to test the impact of photo-$z$ errors on  connectivity. We found that  connectivity is systematically underestimated when using photo-$z$ with respect to exact redshifts. Conversely, the distribution of the connectivity in COSMOS and in the virtual mock photometric {\sc Horizon-AGN} catalogue agree  well;
\item[(3)] Group mass and BGG mass increase with increasing group connectivity. This result is measured both in COSMOS and in {\sc Horizon-AGN}, and is in agreement with our current model of structure formation; 
\item[(4)] At fixed group mass, the mass assembly of the BGG also depends on connectivity. Star-forming BGG are only found in groups of low and medium connectivity. At high group mass, groups with higher connectivity have a lower mass BGG than their counterpart with lower connectivity. 
A consistent result  is found  in {\sc Horizon-AGN}. In the simulation, AGN feedback quenching efficiency is  higher at higher connectivity. Groups in the highest connectivity bin had on average a major merger more recently than at lower connectivity. Alltogether, these findings suggest  that different connectivity values trace different histories of group assembly;
\item[(5)] Given the  correlation between group center and projected skeleton nodes, our finding suggest that 
nodes of the 3D skeleton can also be used to predict  the loci of X-ray groups, as originally suggested by \cite{sousbie112} for clusters.
\end{itemize}
These findings are qualitatively consistent both in  simulations and observational data \citep[see also][for consistent measurements  in the CFHTLS]{sarron19}.
They underline the role played by the large-scale environment (as quantified by connectivity) in shaping a diversity of BGG properties. 
More observations will be needed to make this statement more statistically significant.

In closing, this study is  a first step towards understanding the impact  of the large-scale environment on group mass assembly. 
In order to know precisely what is the physical process beyond the measured correlation between connectivity and BGG mass and type, more detailed measurements from hydrodynamical simulations are essential. 
 Upcoming  3D spectroscopic redshift surveys, including
 4MOST \citep{4MOST}, DESI \citep{DESI}, PFS \citep{PFS}, MSE \citep{2016arXiv160600043M}, provided they allow to extract reliable group catalogues, and  photometric redshifts  surveys including  DES \citep{DES2016}, \emph{Euclid} \citep{Euclid}, WFIRST \citep{WFIRST}, LSST \citep{LSST}, KiDs \citep{KIDS} will
 also prove crucial to extending this pilot study on larger sample and confirming the impact of large-scale environment on group mass assembly. 

Beyond the implication for galaxy formation, cosmology would also benefit from group connectivity measurement as demonstrated by \citet{codis18}. Although it is not possible to make a strong statement on the Dark Energy equation of state from the COSMOS data, the large-scale coverage of \emph{Euclid} and LSST should allow  such statistics.

\section*{Acknowledgments}
{\sl 
EDF was supported by the Oxford physics summer student program when completing this work.
CL is  supported by a Beecroft Fellowship. 
JD  acknowledges funding support from Adrian Beecroft and the STFC. CL would like to thank Oliver Hahn for very relevant suggestions, Yiqing Liu and Hyunmi Song for useful discussions and advices, and the Korean Astronomy and Space science Institute for hospitality when most of this work was finalised. 
CP thanks Sandrine Codis for advices  and  the Royal Observatory in Edinburgh for hospitality when this work was initiated.
This work  relied on the HPC resources of CINES (Jade) under the allocation 2013047012 and c2014047012 made by GENCI
and on the HORIZON and CANDIDE clusters hosted by Institut d'Astrophysique de Paris. We  warmly thank S.~Rouberol for running  the  cluster on which the simulation was  post-processed. 
  This research  is also partly supported by the Centre National d'Etudes Spatiales (CNES) and by the National Science Foundation under Grant No. NSF PHY-1748958. This work is based on data products from
observations made with ESO Telescopes at the La Silla Paranal Observatory under ESO programme ID 179.A-2005 and on data products
produced by TERAPIX and the Cambridge Astronomy Survey Unit on behalf of the UltraVISTA consortium.
We thank D.~Munro for freely distributing his {\sc \small  Yorick} programming language and opengl interface (available at \url{http://yorick.sourceforge.net/}).  
}
\vspace{-0.5cm}

\bibliography{author}

\appendix

\subsection{Impact of  radius choice  to measure  connectivity}
\begin{figure*}
\begin{center}
   \includegraphics[scale=0.55,trim={0.3cm 0 1cm 1.3cm},clip]{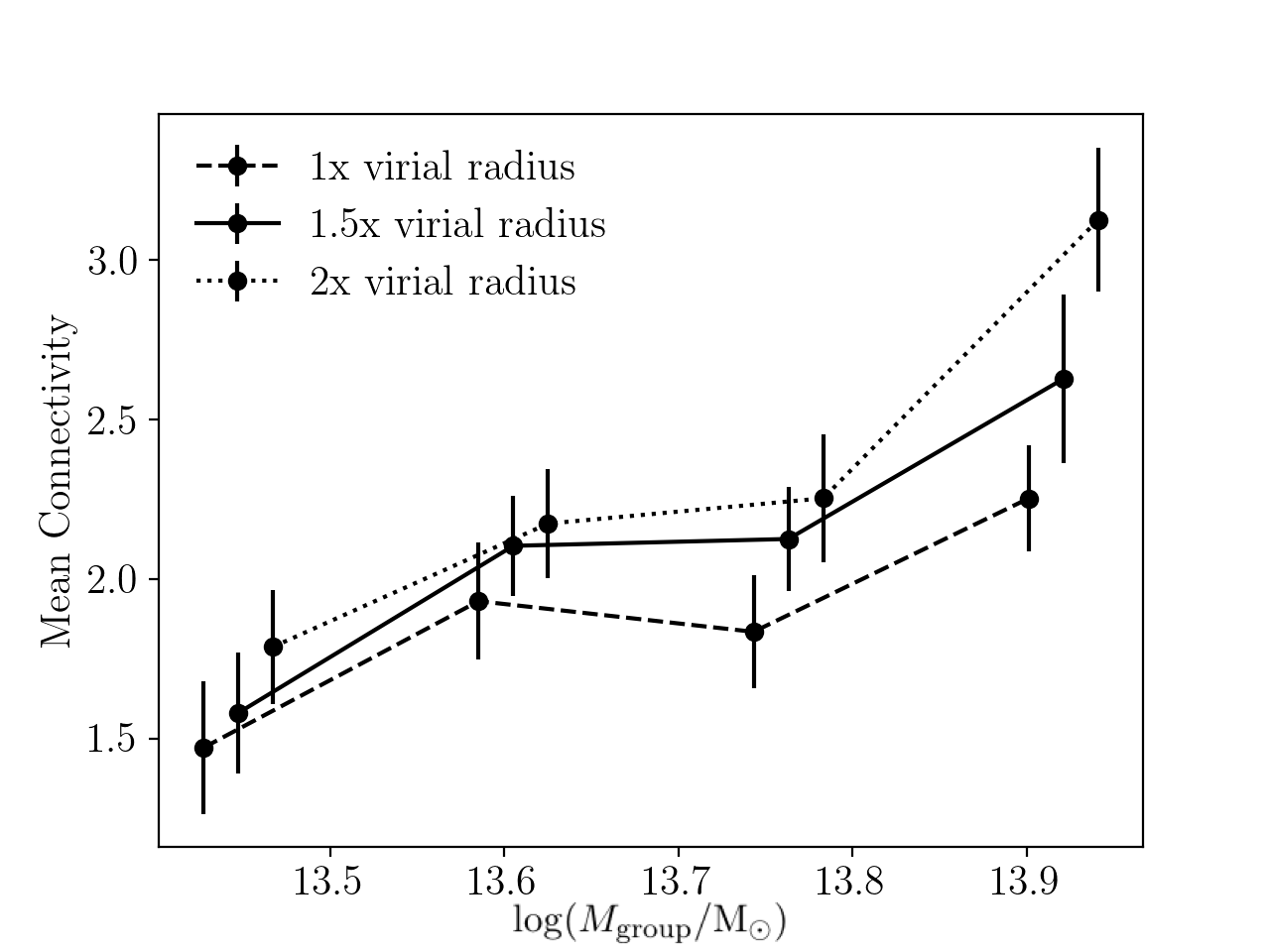}
   \includegraphics[scale=0.55,trim={0.3cm 0 1cm 1.3cm},clip]{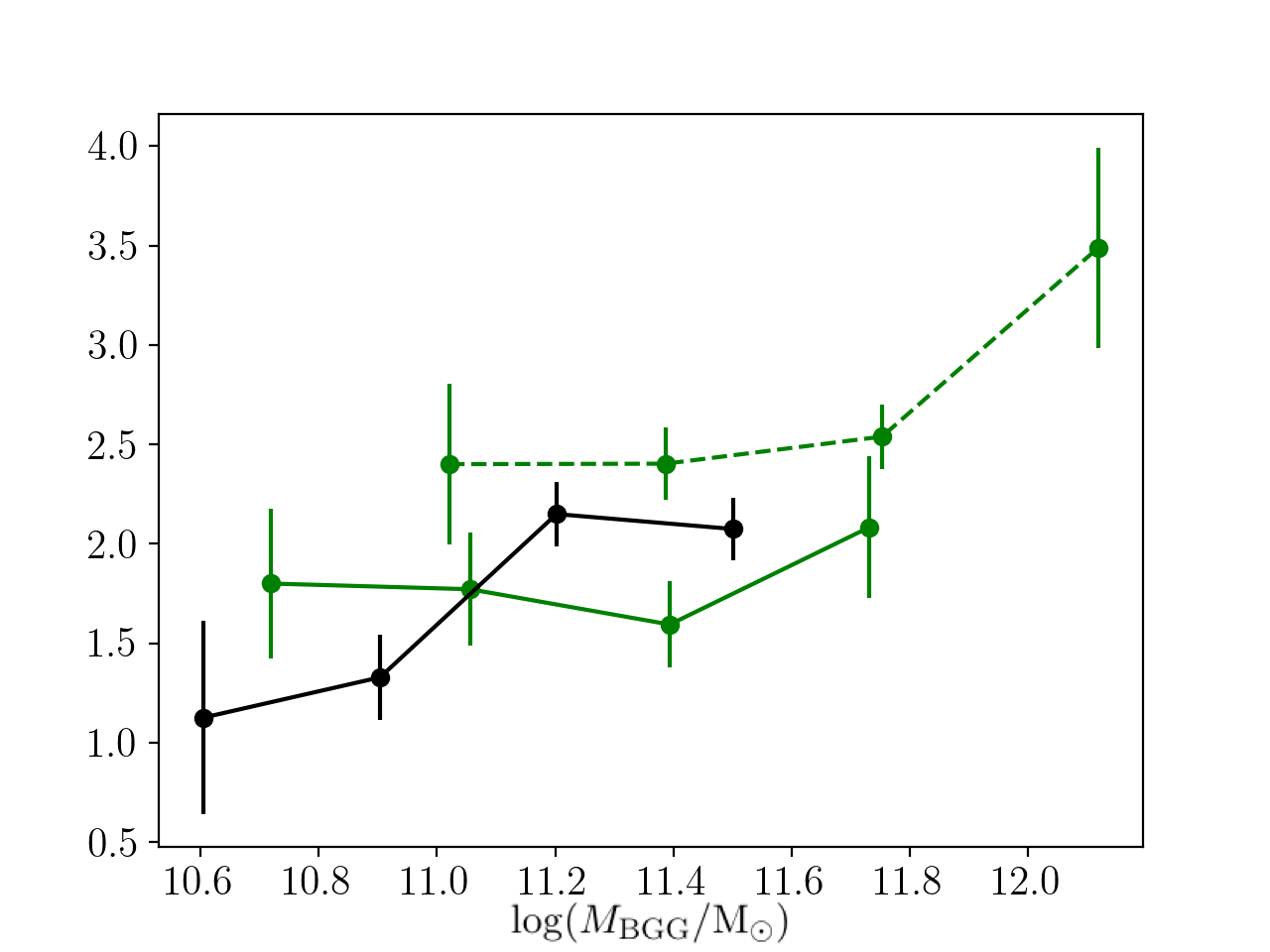}
\end{center}
  \caption{\textit{Left:} Mean connectivity as a function of group mass, for different radii to measure connectivity in $\mathcal{C}_{\rm cosmos}$. \textit{Right}: Mean connectivity as a function of BGG mass when counting type-1b connectivity as connectivity 1 (instead of connectivity 2 as done in the main text) for $\mathcal{C}_{\rm cosmos}$ (\textit{black} line), $\mathcal{C}_{\rm Hzagn 2D}^{\rm phot}$ (\textit{solid green} line), and $\mathcal{C}_{\rm Hzagn 2D}^{\rm phot}$ (\textit{dashed green} line).}
     \label{Fig:MassConnecc}
  \centering

\end{figure*}

We check that our results do not depend on the chosen radius to measure the connectivity. The \textit{left} panel of Fig.~\ref{Fig:MassConnecc} presents the mean connectivity as a function of group mass when measuring the connectivity within 1, 1.5 or 2$\times$ the virial radius of the group. The solid line corresponds to the measurement chosen in the main text. This Figure illustrates that the measured trend of increasing connectivity with group mass does not strongly depend on the chosen radius. 

\section{Counting filaments in COSMOS}
\label{sec:counts}

\subsection{Connectivity 1 versus 2}
\label{App:v1v2}
As discussed in Section~\ref{Sec:photoz}, groups embedded in a filament which do not sit at a peak of the density field (i.e. which are not sitting on a node of the cosmic web) can in principle  be assigned a connectivity of either $\pazocal{C}=1$ (1a$-\pazocal{C}$) or $\pazocal{C}=2$. However we also found that photo-$z$ uncertainties tend to decrease connectivity. In other words, the fact that the group does not sit at a peak of the density field might just be due to shot noise driven by photo-$z$ uncertainties. This is confirmed by the fact that  1a$-\pazocal{C}$ is not found when measuring the connectivity from the galaxy distribution with exact redshift. This fact has driven our choice to consider 1a$-\pazocal{C}$  as $2-\pazocal{C}$ in the subsequent measurements. \\
The \textit{right} panel of Fig.~\ref{Fig:MassConnecc} presents the measurement of mean connectivity versus BGG mass when considering  1a$-\pazocal{C}$  as $1-\pazocal{C}$  (instead of $2-\pazocal{C}$, as was done in the \textit{right} panel of Figure~\ref{Fig:MassConnec}). In this case, the strength of the signal (increase of connectivity for higher BGG masses) decreases when using photo-$z$ (\textit{solid} line) while it is unchanged when using the intrinsic redshift (\textit{dashed} line).

\end{document}